\documentclass[aps,prb,twocolumn]{revtex4-1}

\usepackage{amsmath,amssymb} 
\usepackage{physics} 
\usepackage{graphicx}
\usepackage{dcolumn}
\usepackage{bm}
\usepackage{ulem}
\newcommand\numberthis{\addtocounter{equation}{1}\tag{\theequation}}

\newcommand{\EqLabel}[1]{\label{#1}}

\usepackage{color}
\definecolor{blu}{rgb}{0,0,1}

\definecolor{rd}{rgb}{1,0,0}

\begin{document}


\title{Peierls versus Holstein models for describing electron-phonon coupling in perovskites}


\author{Yau-Chuen Yam}
\author{Mirko M. Moeller}
\author{George A. Sawatzky}%
\author{Mona Berciu}
\affiliation{\!Department \!of \!Physics and Astronomy, \!University
  of\!  British Columbia, \!Vancouver, British \!Columbia,\! Canada,\!
  V6T \!1Z1}
\affiliation{\!Stewart Blusson Quantum Matter \!Institute, \!University
  of British Columbia, \!Vancouver, British \!Columbia, \!Canada,
  \!V6T \!1Z4}

\date{\today}

\begin{abstract}
We use the Momentum Average approximation together with perturbative approaches, in the appropriate limits, to study the single polaron physics on a perovskite lattice inspired by BaBiO$_3$. We investigate electron-phonon coupling of the Peierls type whereby the motion of ions modulates the values of the hopping integrals between sites, and show that it cannot be mapped onto the simpler one-band Holstein model in the whole parameter space. This is because the dispersion of the Peierls polaron has sharp transitions where the ground-state momentum jumps between high-symmetry points in the Brillouin zone, whereas the Holstein polaron always has the same ground-state momentum. These results imply that careful consideration is required to choose the appropriate model for carrier-lattice coupling in such complex lattices.
\end{abstract}

\pacs{Valid PACS appear here}
\maketitle

\section{\label{sec:intro}Introduction}

Materials with perovskite structure ABO$_3$ are known to have a wide
variety of extraordinary properties, ranging from unconventional
high-temperature superconductivity in
cuprates,\cite{gao1994,presland1991,tallon1995,obertelli1992} to an
unusual metal-to-insulator transition in rare-earth nickelates,
\cite{subedi15} to colossal magneto-resistance in
  manganites, \cite{srivastava09,abdelkhalek11}  to multiferroic
  behaviour, \cite{liu2017brief} among others.

Many of these properties are believed to arise from the interplay of
charge, spin, orbital and lattice degrees of freedom and of their
various interactions. A full detailed treatment of all this complexity
is still unfeasible, resulting in the urgent need to identify simpler
but useful models. For instance, is it ever necessary to consider the
full multiplet structure for rare-earths with partially filled $3d$
levels, or does it suffice to include explicitly only one/few of them,
with a simplified description for correlations?
One well-known example
where this kind of question is relevant are the cuprates, where most
models only consider the $3d_{x^2-y^2}$ orbital for
Cu.\cite{anderson1987,emery1987theory,MiJiang20}
Even more basic is the question
of which of the constituent elements need to be included in the
modelling. To continue with the 
CuO$_2$ layer example, even though it is well
known that the doped holes are on the anions, most models do not
explicitly included the oxygen ions. Another example are the
rare-earth nickelates, where only recently it has become clear how
essential it is to include the O explicitly in the
models.\cite{SteveJohsntonPRL2014,AMillis2013} Of course,
the answers will vary from one material to another, but it is
important to ask such questions and to understand when certain
approximations may be valid, and when they are certainly not.

From this perspective, the modeling of the electron-phonon coupling in
perovskites may appear to be on better footing than other issues, as
it has been very customary to use a Holstein coupling to describe it.
\cite{pankaj12,kurdestany17} The Holstein model \cite{holstein1959} is
the simplest possible description of charge carriers interacting with
phonons on a lattice. It was proposed for ``molecular crystals", with
the Einstein mode describing not lattice phonons, but instead an
internal deformation of the individual molecules when an additional
carrier is present. As such, it is not at all obvious that a Holstein
description is appropriate for a complicated system like a perovskite.
Part of the reason for using it is that earlier studies of several
electron-phonon couplings (Holstein,\cite{holstein1959}
Fr\"ohlich,\cite{frohlich1950}, breathing-mode,\cite{kurzynski1976,Bayo2007,Glen2008} etc.) on simple lattices revealed qualitatively
similar behavior \cite{Book}, suggesting that
using the simplest model is likely appropriate. This idea was backed
up for perovskites, under certain assumptions, for a more detailed
model discussed below. Our work challenges this view.

At this point, we must note that an issue that has caused
confusion regarding the importance of electron-phonon
coupling, especially in perovskite structures like the nickelates, the
high-T$_C$ cuprates, the weakly correlated BaBiO$_3$ and also
other systems, is the weak electron-phonon coupling obtained with
ab-initio methods like the Density Functional Theory (DFT). For
example, electron-phonon coupling in cuprates extracted from DFT was
deemed much too small to generate a strong enough pairing for
high-T$_c$ superconductivity.\cite{savrasov1996} However, this is
based on DFT results that predict the $3d_{x^2-y^2}$ bandwidth to be
close to 4 eV, whereas because of the strongly correlated nature of
the Cu $3d$ electrons, the bandwidth of the so-called Zhang-Rice
singlet\cite{zhang1988effective} band is only about 0.3 eV
wide.\cite{yin2008ZRS} This bandwidth renormalization  increases
the effective electron-phonon coupling to the relevant electronic
states by an order of magnitude. This was also noted by Khalliulin and
Horsch in their estimate of a dimensionless
effective coupling $\lambda=0.25$. \cite{khaliullin1997}
At the opposite end of the spectrum, we
note that that interpretation of the ARPES data on
Sr$_2$CuO$_2$Cl$_2$ in terms of polaron formation\cite{shen2004} leads to an
extremely strong effective coupling of $\lambda\approx 6.25$. \cite{Goodvin_2009}

To make the discussion specific, from now on we will use the
perovskite BaBiO$_3$ as our inspiration. This is a good choice because
(i) here there are no complications from strong correlations and/or
spin-orbit coupling (for reasons detailed in the next section), and
(ii) electron-phonon coupling is believed to be strong in this
material, and in fact K-doped BaBiO$_3$ has a record high $T_C\approx 35$K
for a superconductor with a phonon glue. The reason for this high
value of $T_C$ is not yet fully settled. Again, conventional DFT
results predict a much too weak coupling,
\cite{hamada1989,shirai1990,liechtenstein1991,kunc1994,meregalli1998,bazhirov2013}
but a more local molecular-like description of the electronic
structure yields a very substantial electron-phonon coupling, with a
$\lambda \sim 1$.\cite{khazraie2018}

There have already been lots of studies of this very interesting material.
For example, the Rice-Sneddon model described below is widely used
\cite{kostur97,allen97,piekarz00,bischofs02} to investigate the
formation and properties of polarons and bipolarons in BaBiO$_3$. It
has been argued that it maps accurately onto an effective single band
Holstein model,\cite{nourafkan12}
which was then used to study the optical properties
 of BaBiO$_3$ and the metal-insulator transition in
hole-doped BaBiO$_3$. \cite{seibold93}

In this work, we challenge this well-established view that a one-band
Holstein model {\it always} provides a good  description
(at least qualitatively, if not
quantitatively) of the electron-phonon coupling in
perovskites. Our
starting point is a Peierls model that explicitly includes the O sites
and takes into consideration the significant modulation of the hopping
between Bi and O, when the lighter O ions vibrate. This type of
coupling is ignored by the Rice-Sneddon model, which instead focusses
on the change of the on-site energy of the carrier when on a Bi ion,
because of the deformation of the O cage surrounding it.

We are unable to provide accurate results in the physically relevant  limit of
half-filling, {\it i.e.} when there is one hole per unit cell.
Instead, we consider the extreme case when there is a single hole in
the entire system (effectively zero carrier concentration, for an
infinite lattice), because here we can study the properties of the
resulting polaron sufficiently accurately to draw unequivocal
conclusions. Moreover, we investigate the behavior of our model in the
wider parameter space, including regions that are far from where
BaBiO$_3$ is expected to be located. This is partially due to
technical reasons, as the variational approximation that we employ
becomes more accurate for phonon frequencies similar to, or larger
than the electronic intersite hopping integral. As we will show, the
behavior of interest to us appears to evolve smoothly with decreasing
phonon frequencies, so some inferrences can be made about what happens
in the adiabatic limit. Nevertheless, it is important to find other methods
that are reliable in this limit, to verify our conclusions there. A
separate reason to study models with relatively narrow bandwidths, is
the concerted effort in modern condensed matter physics to develop
so-called ``flat-band'' materials, like twisted graphene or ordered
impurity-based midgap bands in semiconductors or insulators. In such
materials, the effective bandwidths can be comparable to or smaller
than the phonon frequencies, and our results would be directly
relevant to them.

Our results demonstrate that in certain regions of the parameter
space, the Peierls model on a perovskite lattice exhibits single polaron
behavior that is impossible to reproduce with a Holstein model.
Specifically, as the electron-phonon coupling is increased, the
polaron dispersion changes its shape such that the ground-state
momentum switches from its free-carrier value to another high-symmetry
point in the Brillouin zone. Such sharp transitions are impossible for
a Holstein polaron.\cite{gerlach91}

Based on this result, we conclude that the Peierls model cannot be
automatically replaced with a Holstein model when studying a
perovskite system.

This being said, it is important to emphasize the caveat that our
study is in the single-polaron limit. It is possible that at finite
carrier concentration, the mapping between Peierls and Holstein couplings might
be valid for some other reasons -- however, this has to be explicitly
verified and not just assumed. To the best of our knowledge, there is
no work addressing this question. We also emphasize that
there are regions of the parameter space (including
  the region where BaBiO$_3$ is believed to be located) where the polaron
ground-state momentum equals the free-carrier value, and therefore a
Holstein model may be sufficient to mimic the polaron behavior for a
correct choice of effective parameters. Again, our point is that this
is not automatically the case for all perovskite materials, therefore
care is needed and one must do detailed work to  justify the
use of a Holstein model as a reasonable description of the
electron-phonon coupling.

The paper is organized as follows: in Section II we introduce our
model, in Section III we describe the various methods we used to study
it, and  in Section IV we present the results. Section V contains
the discussion and conclusions. Technical details are relegated to Appendices.

\section{Model \label{secII}}

\begin{figure}[t]
	\centering
	\includegraphics[width=\columnwidth]{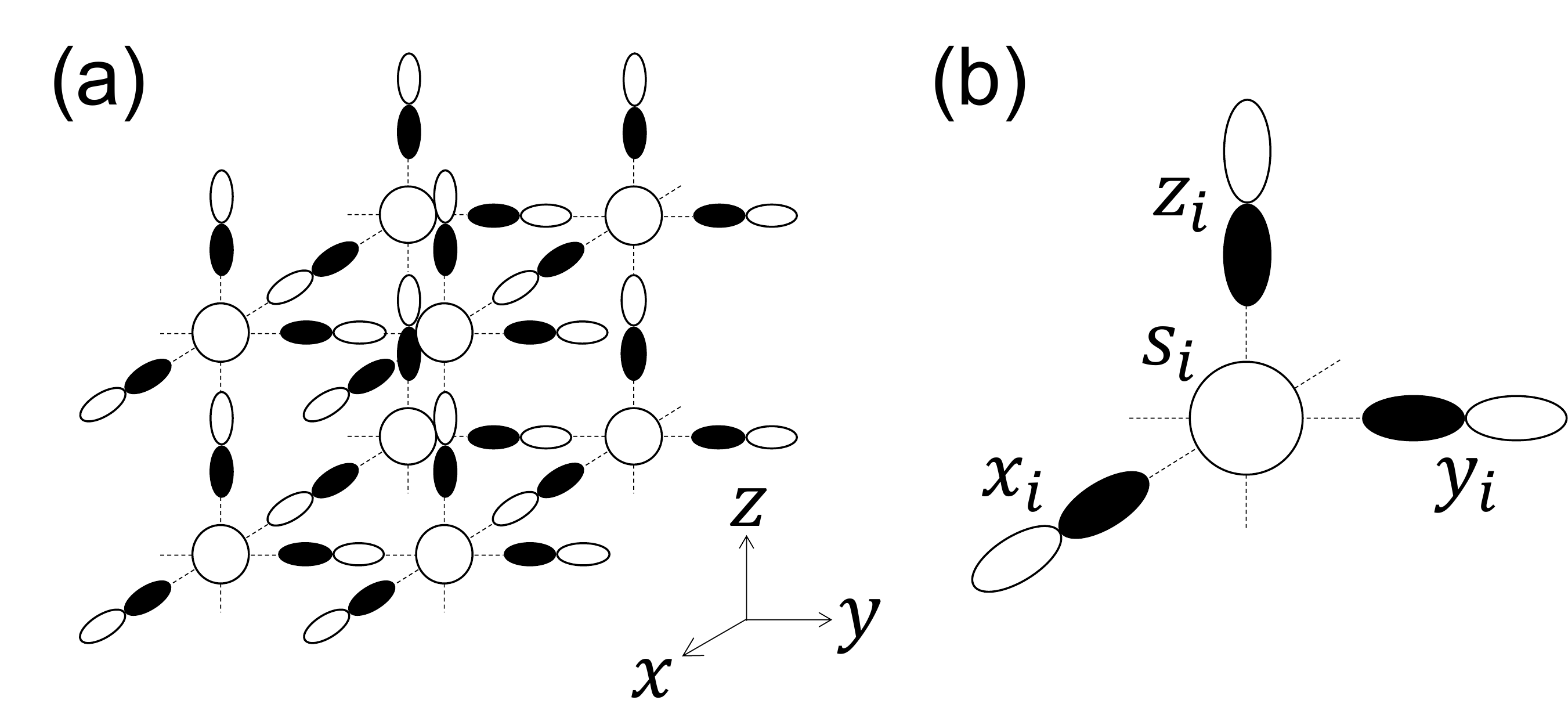}
	\caption{\label{fig1}(a) Sketch of the  model for an infinite 3D lattice, showing $s$ orbitals at the B sites, and the ligand $2p$ orbitals at the O sites. The A sites are ignored. (b) Our choice for the unit cell $i$  has an $s$-orbital labeled by $s_i$ and three $2p$-orbitals labeled $x_i$, $y_i$ and $z_i$ along the three ligand bonds.}
\end{figure}

We use the following approximations to model the generic perovskite ABO$_3$:

(i) sites A are taken to be irrelevant for the physics of interest to us, and are ignored. Physically, this implies that electronic bands with dominant A-character are lying well below and/or well above the Fermi energy. For  BaBiO$_3$, which is our main inspiration, this is a good approximation. It reduces the lattice of interest from a full perovskite to the BO$_3$ lattice sketched in Fig. \ref{fig1}a.

(ii) For the B sites, the relevant electronic orbital is non-degenerate and  spatially well spread out, so that the on-site Hubbard repulsion can be safely ignored. This is a good approximation for BaBiO$_3$, where this is the Bi:$6s$ orbital. From now we will call this the ``s''-orbital, and denote by $s_{i,\sigma}^\dagger$ the creation  operator for a hole with spin $\sigma$ in this orbital of the atom B  in the unit cell $i$. 

(iii) At each O site, we only keep in the model the $2p_{\gamma}$ orbital with ligand character, {\it i.e.} $\gamma = x, y, z$ for the O located on bonds parallel to $x,y,z$, respectively. We will refer to this as the $x,y$ or $z$ orbital, and use either the generic $\gamma^\dagger_{i,\sigma}$ operator when refering to any of the three O in the unit cell $i$, or the specific $x^\dagger_{i,\sigma}, y^\dagger_{i,\sigma},  z^\dagger_{i,\sigma}$ for the creation operator associated with adding a  hole to the O located on the $x,y,z$ bond of unit cell $i$, see Fig. \ref{fig1}(b). 

(iv) We ignore all phonon modes that are primarily located on A and B sites, and instead keep only the optical phonon describing longitudinal (parallel to its ligand bond) oscillations of each O. The first part is reasonable as the A and B atoms are much heavier than O, so we expect their motion to mostly contribute to very low-energy phonon modes which do not couple strongly to the hole's motion (see below). The second part is justified because to first order, one can think of each O as oscillating longitudinally between its two immobile B neighbours, with a characteristic frequency $\Omega$ that is the same at all O sites. For a crystal, this is equivalent with an Einstein phonon mode of frequency $\Omega$ on the O sites. In the following, we will denote the phonon creation operator for the $\gamma=\{x,y,z\}$ O site in unit cell $i$ as $b^\dagger_{i,\gamma}$.

In this work, we focus on the effect of this phonon mode on the hybridization  between neighbor O and  B sites, as well as neighbor O sites. The resulting electron-phonon coupling is known as a Peierls coupling, and should be contrasted to the Rice-Sneddon model that focuses on the modulation of the on-site energy of a hole located in the $s$-orbital, due to oscillatory motion of adjacent O. As discussed in the Introduction, the latter has been argued to be well modelled by an effective Holstein coupling on a simplified cubic lattice with only $B$ sites included. Our results discussed below show that   this equivalence with a Holstein model does not hold for the Peierls coupling in a considerable region of the parameter space.

To summarize, the Peierls model that we study is: 
\begin{equation}
\EqLabel{H}
\hat{H} =\hat{H_0}+\hat{V},
\end{equation}
where
\begin{equation}
\EqLabel{Ho}
\hat{H_0}=\Omega\sum_{i, \gamma}b^{\dagger}_{i\gamma}b_{i\gamma}-\Delta\sum_{i,\gamma} \gamma^{\dagger}_i\gamma_i
+T_{sp}+T_{pp}
\end{equation}
describes the Einstein phonon modes (we set $\hbar = 1$), the charge-transfer energy  $\Delta$ between $p$ and $s$ atomic orbitals,  and the nearest neighbor (nn) $s$-$p$ and $p$-$p$ hopping, respectively, while
\begin{equation}
\EqLabel{V}
\hat{V}=H^{sp}+H^{pp}
\end{equation}
is the Peierls electron-phonon coupling describing the modulation of the $s$-$p$ and $p$-$p$ hoppings due to the O vibrations. Specifically:
\begin{align*}
&T_{sp}=t\sum_{i,\gamma}s^{\dagger}_i(\gamma_i-\gamma_{i-\gamma})+h.c.\\
&T_{pp}=-t_p\sum_{i,\gamma}\gamma^{\dagger}_i
(\gamma'_{i}-\gamma'_{i-\gamma'}-\gamma'_{i+\gamma}+\gamma'_{i+\gamma-\gamma'})
+h.c.\\
&H^{sp}=-\alpha t\sum_{i,\gamma}[\gamma^{\dagger}_i
(s_i+s_{i+\gamma})(b^{\dagger}_{i\gamma}+b_{i\gamma})+h.c.]\\
&\begin{aligned}
H^{pp}=&\beta t_p
\sum_{i,\gamma}
[\gamma^{\dagger}_i(\gamma'_{i}-\gamma'_{i-\gamma'}+\gamma'_{i+\gamma}-\gamma'_{i+\gamma-\gamma'}\\
&+\gamma''_{i}-\gamma''_{i-\gamma''}+\gamma''_{i+\gamma}-\gamma''_{i+\gamma-\gamma''})
(b^{\dagger}_{i\gamma}+b_{i\gamma})+h.c.]
\end{aligned}
\end{align*}
where we use the short-hand notation:
\begin{align*}
\gamma'=\left\{ \begin{array}{rcl}
y, & \mbox{if } \gamma=x \\
z, & \mbox{if } \gamma=y \\
x, &\mbox{if }  \gamma=z 
\end{array}\right.; \quad
\gamma''=\left\{ \begin{array}{rcl}
z, & \mbox{if } \gamma=x \\
x, & \mbox{if } \gamma=y \\
y, & \mbox{if } \gamma=z 
\end{array}\right.
\end{align*}
in the above sums.

We note that here and in the following we ignore the spin degree of freedom $\sigma$ of the hole, which is irrelevant in the one-hole limit we study below.

Apart from $\Omega$, the parameters are the charge-transfer energy $\Delta$ and the hopping integrals $t$ and $t_p$ for $s$-$p$ and $p$-$p$ hopping, respectively, when the O are at their equilibrium positions. The latter are  negative numbers $t,t_p<0$ for holes, with the additional signs due to the orbitals' overlaps explicitly written in the Hamiltonians above. Similarly, $\alpha$ and $\beta$ characterize the electron-phonon couplings coming from the modulation of the $s$-$p$ and $p$-$p$ hoppings when the O are displaced out of their equilibrium positions. For holes, $\alpha,\beta>0$ and according to  Harrison's rule, $\beta=\alpha/2$.\cite{harrison}

\subsection{The BO$_6$ cluster model}

Density functional theory (DFT) studies  of BaBiO$_3$ revealed that the most important hybridization is between the $s$ orbital and the linear combination of neighbor O $p$-orbitals with A$_{1g}$ symmetry. \cite{foyevtsova15} This stabilizes ``molecular"-like  orbitals with  $s+ p_{A_{1g}}$ character, and suggests a possible mapping onto a simple cubic lattice, by retaining only the lowest such state for each BO$_6$ cluster.

\begin{figure}[t]
	\centering
	\includegraphics[width=0.8\columnwidth]{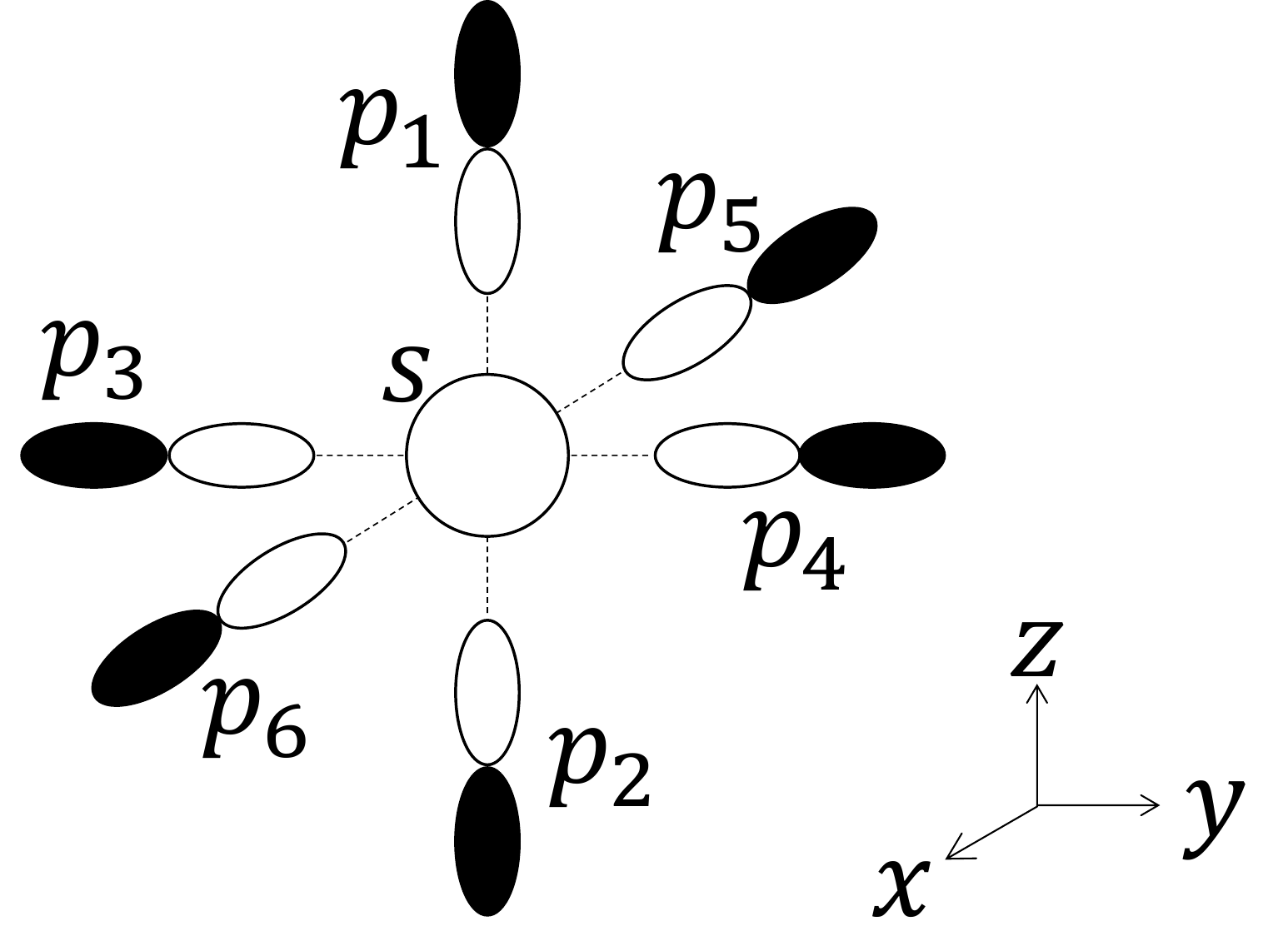}
	\caption{\label{fig2} BO$_6$ cluster with the central $s$ orbital surrounded by six $p$ ligand orbitals. Note that for convenience, here we use a different convention for the signs of the {$p_1, p_4$ and $p_6$} orbitals than used in the lattice case depicted in Fig. \ref{fig1}.}
\end{figure}

To test this hypothesis, we  also investigate a single BO$_6$ cluster and the effects of Peierls coupling on its spectrum. The Hamiltonian is that of Eq. (\ref{H}) when limited to a single $B$ site and its 6 O neighbours. For convenience, for the cluster case we choose a different convention for the signs of the $2p$ orbitals' lobes, as shown in Fig. \ref{fig2}.

The corresponding cluster model is:
\begin{align*}
	\mathcal{H} = &\Omega\sum^6_{i=1}b^{\dagger}_i b_i+\Delta s^{\dagger}s  -
	\sum^6_{i=1} t(s^{\dagger}p_i+p_i^{\dagger}s)\\
	&-\sum^6_{i=1}\alpha t(s^{\dagger}p_i+p^{\dagger}_is)(b^{\dagger}_i+b_i)\\
	&-t_p\sum_{i=1,3,5}\left[(p^{\dagger}_i+p^{\dagger}_{i+1})
	(p_{i+2}+p_{i+3})
	+\text{h.c.}\right]\\
	&-\beta t_p\big[[p^{\dagger}_6(b^{\dagger}_6+b_6)+
	p^{\dagger}_5(b^{\dagger}_5+b_5)](p_4+p_3)\\
	&+(p^{\dagger}_5+p^{\dagger}_6)
	[p_3(b^{\dagger}_3+b_3)+p_4(b^{\dagger}_4+b_4)]
	+\dots\big]
	\numberthis \label{H_cluster}
\end{align*}
where $i$ labels are cyclic with period 6. Note that we only listed explicitly a few of the terms in $H^{pp}$ (the last two lines). Writing all of them makes the equation too long, and is not  illuminating.   All the parameters have the same meaning as in Hamiltonian (\ref{H}).

    The cluster Hamiltonian can be written more simply in terms of hole and boson operators consistent with its symmetry. We define the new hole operators:
\begin{align*}
P_1&=\frac{1}{\sqrt{6}}(p_1+p_2+\dots+p_6)\\
P_2&=\frac{1}{\sqrt{12}}(2p_1+2p_2-p_3-p_4-p_5-p_6)\\
P_3&=\frac{1}{\sqrt{4}}(p_5+p_6-p_3-p_4)\\
P_4&=\frac{1}{\sqrt{2}}(p_6-p_5)\\
P_5&=\frac{1}{\sqrt{2}}(p_4-p_3)\\
P_6&=\frac{1}{\sqrt{2}}(p_1-p_2)
\end{align*}
and similarly for the boson operators: $B_1 = \sum_{i=1}^{6}b_i$, etc. 

$P^\dagger_1$ creates a hole in the linear combination of O $2p$ orbitals with $A_{1g}$ ($s$-like) symmetry, $P^\dagger_2$ and $P^\dagger_3$ correspond to the $E_g$ terms with $d_{3z^2-r^2}$ and $d_{x^2-y^2}$ symmetry, respectively,  and $P^\dagger_4, P^\dagger_5$ and $P^\dagger_6$ correspond to the $T_{1u}$ terms with $p_x$, $p_y$ and $p_z$ symmetry, respectively. We define the new bosonic operators $B_i$ similarly.

In the new basis, the cluster Hamiltonian becomes:
\begin{align*}
	&\mathcal{H}=\Omega\sum_{i=1}^{6}B_i^{\dagger}B_i+ \Delta s^{\dagger}s
	-T(s^{\dagger}P_1+P_1^{\dagger}s)\\
	&-\alpha t\sum^6_{i=1}(s^{\dagger}P_i+P_i^{\dagger}s)(B^{\dagger}_i+B_i)\\
        & - t_p(4P_1^{\dagger}P_1-2P_2^{\dagger}P_2
-2P_3^{\dagger}P_3)\\
&- \beta t_p\bigg(\sqrt{\frac{8}{3}}(B_1^{\dagger}+B_1)
(2P_1^{\dagger}P_1-P_2^{\dagger}P_2-P_3^{\dagger}P_3) +\dots 
\bigg)\numberthis\label{H_sym}
\end{align*}
where the dots in the last line refer to terms involving phonon operators $B_i+B_i^\dagger$ with $i\ne 1$, \textit{i.e.} the other $E_g$ and $T_{1u}$ symmetries are included. They are straightforward but too lengthy to write here. Note that because of the $pp$ hopping, the effective charge transfer energy between the $s$ and the $P_1$ molecular orbital is $\Delta_1=\Delta+4t_p$, whereas for the $E_g$ molecular orbitals, the effective charge transfer energy is  $\Delta_2=\Delta_3=\Delta+2t_p$. 

The third term shows that indeed, the $s$ orbital only hybridizes with the $P_1$ molecular orbital with the same $A_{1g}$ symmetry, and the  effective hopping is $T=\sqrt{6} t$. However, because the presence of deformations breaks this symmetry, the Peierls  $\alpha$ electron-phonon coupling allows hopping between the $s$ and any of the $P_i$ orbitals, if bosons with the same symmetry $i$ are either already present, or are being created during the process -- see terms on the second line. Terms due to the $p$-$p$ hopping can be understood similarly.

\section{Methods \label{secIII}}

We studied the models described above by a variety of means which we briefly review here, with full details relegated to various appendixes.

\subsection{Perturbation theory for the lattice case in the anti-adiabatic limit}

In the antiadiabatic limit where $\Omega$ is the largest energy scale, we can use perturbation theory to project out the high-energy states with one or more phonons, to obtain an effective Hamiltonian describing the motion of the polaron. The resulting analytical dispersion is useful  because it allows us to gain intuition about the behavior of the polaron in this limit, as discussed below.

We partition the Hamiltonian into $\hat{H}=\hat{h}_0+ \hat{h}_1$, where $\hat{h}_0\equiv \Omega\sum_{i,\gamma} b^{\dagger}_{i\gamma}b_{i\gamma}$ is the large part, and $\hat{h}_1$ includes all the other terms and is treated as the perturbation. Using standard second order perturbation theory (PT),\cite{Takahashi} we obtain the low-energy effective Hamiltonian to be:
\begin{align*}
\hat{h}=\hat{h}_0+\hat{P_0}\hat{h}_1\hat{P_0}
+\hat{P_0}\hat{h}_1
\frac{1-\hat{P_0}}{E_0-\hat{h}_0}\hat{h}_1\hat{P_0}
+{\cal O}\left({1\over \Omega^2}\right)
\end{align*}
where $\hat{P_0}$ is the projection operator onto the  highly-degenerate, one-hole ground state manifold of $\hat{h}_0$, {\it i.e.} zero-phonon states with energy $E_0=0$. 

After carrying out these calculations, we find that $\hat{h}=-\Delta \sum_{i\gamma}^{} \gamma^\dagger_i \gamma_i + T_{sp}+T_{pp} + \delta\hat{h} +{\cal O}\left({1\over \Omega^2}\right) $ where:
\begin{align}
\delta\hat{h}=-\frac{\alpha^2t^2}{\Omega}\sum_{j\gamma}
 (\tilde{s}^{\dagger}_{j,\gamma}+\tilde{s}^{\dagger}_{j-\gamma,\gamma})s_j
 - \frac{2\alpha^2t^2+8\beta^2 t_p^2}{\Omega}\sum_{j\gamma}^{} \gamma^{\dagger}_j\gamma_j&
\nonumber \\ 
+\frac{\alpha\beta tt_p}{\Omega}\sum_{j\gamma}(\bar{\gamma}^{\dagger}_{j,\gamma'}
+\bar{\gamma}^{\dagger}_{j,\gamma''}
+\bar{\gamma}^{\dagger}_{j-\gamma,\gamma'}
+\bar{\gamma}^{\dagger}_{j-\gamma,\gamma''})s_j \qquad &\nonumber \\
 -\frac{\alpha \beta t t_p }{\Omega}\sum_{j\gamma} (
-\tilde{s}^{\dagger}_{j,\gamma''}+
\tilde{s}^{\dagger}_{j+\gamma,\gamma''}-
\tilde{s}^{\dagger}_{j-\gamma'',\gamma''}+
\tilde{s}^{\dagger}_{j+\gamma-\gamma'',\gamma''}&\nonumber \\
 -\tilde{s}^{\dagger}_{j,\gamma'}
+\tilde{s}^{\dagger}_{j+\gamma,\gamma'}
-\tilde{s}^{\dagger}_{j-\gamma',\gamma'}
+\tilde{s}^{\dagger}_{j+\gamma-\gamma',\gamma'})\gamma_j\qquad &\nonumber \\
- \frac{\beta^2 t_p^2}{\Omega} \sum_{j\gamma}(\bar{\gamma}''^{\dagger}_{j,\gamma}
+\bar{\gamma}''^{\dagger}_{j,\gamma'}
-\bar{\gamma}''^{\dagger}_{j+\gamma,\gamma}
-\bar{\gamma}''^{\dagger}_{j+\gamma,\gamma'}
+\bar{\gamma}''^{\dagger}_{j-\gamma'',\gamma}\qquad & \nonumber \\
+\bar{\gamma}''^{\dagger}_{j-\gamma'',\gamma'}
 -\bar{\gamma}''^{\dagger}_{j+\gamma-\gamma'',\gamma}
-\bar{\gamma}''^{\dagger}_{j+\gamma-\gamma'',\gamma'}
 +\bar{\gamma}'^{\dagger}_{j,\gamma''}\qquad &\nonumber \\
+\bar{\gamma}'^{\dagger}_{j,\gamma}
 -\bar{\gamma}'^{\dagger}_{j+\gamma,\gamma''}
-\bar{\gamma}'^{\dagger}_{j+\gamma,\gamma}
+\bar{\gamma}'^{\dagger}_{j-\gamma',\gamma''}\qquad &\nonumber \\
\label{delta_h}
+\bar{\gamma}'^{\dagger}_{j-\gamma',\gamma}
-\bar{\gamma}'^{\dagger}_{j+\gamma-\gamma',\gamma''}
-\bar{\gamma}'^{\dagger}_{j+\gamma-\gamma',\gamma})\gamma_j\qquad &
\end{align}
and we used the short-hand notation:
\begin{align*}
\tilde{s}^{\dagger}_{j,\gamma}&\equiv s^{\dagger}_j+s^{\dagger}_{j+\gamma}\\
\bar{\gamma}^{\dagger}_{j,\gamma'}&\equiv
\gamma'^{\dagger}_j
-\gamma'^{\dagger}_{j-\gamma'}
+\gamma'^{\dagger}_{j+\gamma}
-\gamma'^{\dagger}_{j+\gamma-\gamma'}.
\end{align*}

The expression of $\delta\hat{h}$ may seem complicated, but it consists of simple terms whose appearance is conceptually straightforward to understand. They can be divided into on-site energies like $-6\alpha^2t^2/\Omega\sum_{j}s^\dagger_js_j$ (part of the first term on the first line), which reflect the polaron formation energy as a hole located at an $s$-site hops to a neighbor O and back while creating and then reabsorbing a phonon at that O site. The on-site energy at the O sites is also renormalized (last term on the first line) but by a different amount, so together  these two terms imply a change of the efective $\Delta$.

All other terms describe longer-range hoping  {\it dynamically generated} through phonon emission+absorption. For example, the first term on the first line contains terms proportional to $s^\dagger_{j'} s_j$, where $j'$ and $j$ are nn neighbor $s$ orbitals. These terms are generated when a hole hops from site $j$ to the O located in between $j$ and $j'$ while creating a phonon at that O, and then hops again while absorbing the phonon, and lands at site $j'$. Similar processes generate additional $s$-$p$ and $p$-$p$ hoppings, which supplement and renormalize the bare hopping $ T_{sp}+T_{pp} $ and will therefore modify the polaron dispersion. 

To find the polaron dispersion, we Fourier transform  $\hat{h}$. For any $\bm{k}$-point in the cubic Brillouin zone, we get a $4\times 4$ matrix that can be diagonalized numerically. The lowest of the four bands is the polaron band.

\subsection{Perturbation theory for the lattice case for weak electron-phonon coupling}

Another case that can be treated with standard PT is when the electron-phonon coupling $\alpha \rightarrow 0$. For simplicity, we set $\beta=0$ and treat only the equivalent 1D case\cite{moller2016}  -- this suffices for our needs. The more general 3D case with $\beta \ne 0$ can be treated similarly.

Using Rayleigh-Schr\"{o}dinger perturbation to second-order, the polaron energy is:
\begin{widetext}
\begin{align}
E_P(k)=E_0(k) + \frac{(\alpha t)^2}{8\pi}\int_{-\pi}^{\pi} dq  \bigg[&-(1+e^{-i(k-q)})(1+e^{ik})
\left(\frac{1}{E_0(k)-\Omega-E_0(k-q)}-\frac{1}{E_0(k)-\Omega+E_0(k-q)}\right)\nonumber\\
&+(1+e^{-i(k-q)})(1+e^{i(k-q)})
\left(\frac{1}{E_0(k)-\Omega-E_0(k-q)}+\frac{1}{E_0(k)-\Omega+E_0(k-q)}\right)\nonumber\\
&+(1+e^{ik})(1+e^{-ik})
\left(\frac{1}{E_0(k)-\Omega-E_0(k+q)}+\frac{1}{E_0(k)-\Omega+E_0(k+q)}\right)\nonumber\\
&-(1+e^{-ik})(1+e^{i(k+q)})
\left(\frac{1}{E_0(k)-\Omega-E_0(k+q)}-\frac{1}{E_0(k)-\Omega+E_0(k+q)}\right)
\bigg] \label{wPT}
\end{align}
\end{widetext}
where $E_0(k)=-|2t\sin(k/2)|$ is the free-hole dispersion and $a=1$. This result is only valid for phonon energy $\Omega>2|t|$, because otherwise the denominators vanish for large enough $k$ (Brillouin-Wigner PT must be used in this case). For $\alpha=0$, the GS is at $k=\pi$ and has energy $E_0(\pi)=-2|t|$, while $E_0(0)=0$. Note the additional phase-factors inside the integrand. These appear because the Peierls electron-phonon vertex, when Fourier transformed, depends explicitly on both the hole momentum $k$  and the phonon momentum $q$. This $(k,q)$ dependence is a direct consequence of the non-diagonal nature of the Peierls coupling, and  is very unlike the Holstein model, where this vertex is a constant.

\subsection{Momentum Average (MA) approximation for the lattice case}

MA is a variational method for calculating the one-hole Green's functions
$G^{\beta\alpha}(\bf{k},\omega)\equiv \mel{0}{\beta_{\bf{k}}\hat{G}(\omega)\alpha^{\dagger}_{\bf{k}}}{0}$,
where $\alpha,\beta\in \{s,x,y,z\}$ are any pair of orbitals, $\hat{G}(\omega)=\left[\omega+i\eta-\hat{H}\right]^{-1}$ is the rezolvent for the  Hamiltonian of Eq. (\ref{H}), and $|\bm{0}\rangle$ is the vacuum for holes and phonons.

Here we present a brief overview of MA, with technical details relegated to Appendix \ref{appMA}.
To find $G^{\beta\alpha}(\bf{k},\omega)$, we  use Dyson's identity: $\hat{G}(\omega)=
\hat{G_0}(\omega)+\hat{G}(\omega)\hat{V}\hat{G_0}(\omega)$, where $\hat{H_0}$ is the Hamiltonian of Eq. (\ref{Ho}), $\hat{V} =\hat{H}- \hat{H_0}$, and $\hat{G_0}(\omega)=[\omega+i\eta-\hat{H_0}]^{-1}$ is the rezolvent for $\hat{H_0}$, whose corresponding propagators $G_0^{\beta\alpha}(\bf{k},\omega)=\mel{0}{\beta_{\bf{k}}\hat{G_0}(\omega)\alpha^{\dagger}_{\bf{k}}}{0}$ can be calculated by Chebyshev Polynomials expansion, as explained in  Appendix \ref{CP}. Using Dyson's identity leads to the exact equation:
\begin{widetext}
  \begin{align}
    \label{G}
G^{\beta\alpha}({\bf{k}},\omega) = &G^{\beta\alpha}_0({\bf{k}},\omega)-\alpha t\sum_{\gamma}
(1+e^{i {\bf{k}}_{\gamma} a})\tilde{f}^{(1)}_{\gamma,\gamma} G^{s \alpha}_0 ({\bf{k}},\omega) \nonumber \\
&-\sum_{\gamma}[\alpha t\tilde{f}^{(1)}_{s,\gamma}-
\beta t_p(\bar{f}^{(1)}_{\gamma',\gamma}
+\bar{f}^{(1)}_{\gamma'',\gamma}
 +\xi_{\gamma''\gamma}({\bf{k}})\bar{f}^{(1)}_{\gamma'',\gamma''} +\xi_{\gamma'\gamma}({\bf{k}})\bar{f}^{(1)}_{\gamma',\gamma'})
]G^{\gamma\alpha}_0({\bf{k}}, \omega)
\end{align}
\end{widetext}
where we defined the generalized propagators
\begin{equation}
\EqLabel{fs}
f^{(n)}_{\gamma,\delta,\ell}({\bf{k}},\omega)\equiv
\sum_j \frac{e^{i{\bf{k}}R_j}}{N}\mel{0}{\beta_{\bf{k}}\hat{G}(\omega)
	\gamma^{\dagger}_{j+\ell}(b^{\dagger}_{j\delta})^n}{0}
\end{equation}
and we use the short-hand notations:
\begin{align*}
&\xi_{\gamma_1,\gamma_2}({\bf{k}}) \equiv 1-e^{-i{\bf{k}}_{\gamma_2}a}+e^{i{\bf{k}}_{\gamma_1}a}
-e^{i ({\bf{k}}_{\gamma_1}-{\bf{k}}_{\gamma_2}) a}\\
&\tilde{f}^{(n)}_{s,\gamma} \equiv f^{(n)}_{s,\gamma,0}+f^{(n)}_{s,\gamma,\gamma} \\
&\tilde{f}^{(n)}_{\gamma,\gamma} \equiv f^{(n)}_{\gamma,\gamma,0}\\
&\bar{f}^{(n)}_{\gamma_1,\gamma_2}  \equiv f^{(n)}_{\gamma_1,\gamma_2,0}-f^{(n)}_{\gamma_1,\gamma_2,-\gamma_1}
+f^{(n)}_{\gamma_1,\gamma_2,\gamma_2}-f^{(n)}_{\gamma_1,\gamma_2,\gamma_2-\gamma_1}\\
\end{align*}
in which for simplicity, the dependence on $({\bf{k}},\omega)$ of the various $f$-propagators
is not written explicitly.

To find equations of motion for the various $f^{(1)}$ propagators appearing in Eq. (\ref{G}), we apply again Dyson's identity. The electron-phonon coupling terms either remove  the phonon, linking the  various $f^{(1)}$ back to various $G^{\beta\alpha}({\bf{k}},\omega)$, or add a phonon and thus also link to new propagators with two phonons present in the initial (ket) state. If the two phonons are on the same O site, the corresponding propagator is one of the $f^{(2)}$ defined in Eq. (\ref{fs}) and we keep it, while we ignore the propagators with phonons located on different sites. The same procedure is employed to generate equations of motion for all $f^{(n)}$ for any $n\ge 2$, linking them to various $f^{(n-1)}$ and $f^{(n+1)}$.

The resulting equations, listed in Appendix (\ref{appMA}) where we also discuss their solution, implement the variational guess that the largest weight to the polaron cloud comes from configurations where all phonons are at the same O site. That this should be a reasonable choice can be seen as follows: (i) if the hole is at an O site that is already displaced, {\it i.e.} it has phonons, the Peierls electron-phonon coupling $\alpha$ will hop it to one of its neighbor B sites and create an additional phonon at the original O site - this process is included in our variational calculation. The Peierls electron-phonon coupling $\beta$ will hope the hole to an adjacent O site, creating a new phonon either at the original O site (a process we include), or at the new O site (a process we ignore, because now there would be phonons on two different sites). Similarly, if (ii) the hole is at a B site neighbor to an O with several phonons, then a Peierls $\alpha$ process can take the hole back to the displaced O site, adding to the number of phonons there (we keep this), or to a different O site (we dismiss this as it would add a phonon at the new site). The reason is that each phonon costs an energy $\Omega$ but the hole cannot take advantage of (interact simultaneously with) phonons on multiple sites, so the most advantageous low-energy approach is to keep the phonon cloud spatially small.

We have tested this intuition for the 1D version of this model in Ref. \onlinecite{moller2016}, where we compared the MA results against those of exact diagonalization (ED) with excellent success. ED is prohibitively expensive in higher dimensions, but we know from extensive studies of MA for other models that its accuracy improves with increasing dimensionality \cite{PRL06,PRB06,PRB07}. Additionally, MA accuracy can be gauged by increasing the variational space. The simplest new configurations are those allowing an extra phonon on a different site than the one that already has a cloud. For the 2D version of this model we found that including these additional states has very small influence on the results, for instance changing eigenenergies by very few percent. \cite{moller2017} We have not implemented the expanded variational calculation here because it becomes much more cumbersome in 3D and such small quantitative variations will not affect the conclusions we draw below. 

\subsection{Variational approximation for the cluster Hamiltonian}

The spectrum of the cluster Hamiltonian of Eq. (\ref{H_sym}) can be found by exact diagonalization, but for our purposes it suffices to use a variational approximation that sets an upper bound to the ground-state energy. The best trial wavefunction we found is:
\begin{equation}
\EqLabel{trial}
\ket{\psi_v}=\frac{s^{\dagger}+\sum_{i=1}^{6}\chi_iP_i^{\dagger}}
{\sqrt{1+\sum_{i=1}^{6}\chi_i^2}}\prod_{i=1}^{6} e^{-\frac{1}{2}\eta_i^2+\eta_iB^{\dagger}_i}\ket{0}
\end{equation}
where $\chi_i$ and $\eta_i$ are variational parameters defining the electronic part of the ``cluster" orbital, and the coherent distortions associated with the various symmetries, respectively.

After some algebra, we find:
\begin{equation*}
\EqLabel{Evar}
\mel{\psi_v}{{\cal H}}{\psi_v}=
\frac{\Delta-2t\chi_1}{1+\sum_i{\chi}_i^2}
+\Omega \sum_{_i}^{}\eta_i^2 - \frac{4 {\alpha t} \sum_{i}^{} \chi_i \eta_i}{1+\sum_i{\chi}_i^2} + \dots
\end{equation*}
where the dots are terms coming from $p$-$p$ hopping, which we do not write here so as to keep the expression compact (however, these terms are included in the calculation). This is minimized numerically to find $\chi_i$ and $\eta_i$, which are then used to calculate an upper bound $\mel{\psi_v}{{\cal H}}{\psi_v}$ for the cluster ground-state energy, which we will refer to as the ``variational" cluster energy.

\subsection{$A_{1g}$ approximation for the cluster Hamiltonian}

Given that the $s$ orbital only hybridizes with the  $P_1$ orbital with a large $T=\sqrt{6} t$, one may expect that the terms with  $A_{1g}$ symmetry contribute most to the ground-state. This implies a symmetric, $s$-like distortion of the O cage.  We could then remove from the cluster Hamiltonian the terms with boson operators of other symmetries, and still expect a good low-energy description.

The resulting simplified cluster Hamiltonian is:
\begin{align*}
{\cal H}_{A_{1g}}=&\Delta s^{\dagger}s+\Omega B_1^{\dagger}B_1-T(s^{\dagger}P_1+P_1^{\dagger}s)\\
&-\alpha t(s^{\dagger}P_1+P_1^{\dagger}s)(B_1^{\dagger}+B_1)\\
&-4t_pP_1^{\dagger}P_1-4\sqrt{\frac{2}{3}}\beta t_pP_1^{\dagger}P_1(B_1^{\dagger}+B_1)
\numberthis\label{HA1g}
\end{align*}

We note again that the effective charge transfer energy is $\Delta_1=\Delta+4t_p$, however to make comparisons easier, we continue to work with the original parameters $\Delta$ and $t_p$.

The ground state energy for this Hamiltonian can be found using continued fractions, \cite{PRBR07} see also Appendix \ref{AppC}. We will refer to it as the ``$A_{1g}$" cluster energy.

\begin{figure*}
	\centering
	\includegraphics[width=0.83\paperwidth]{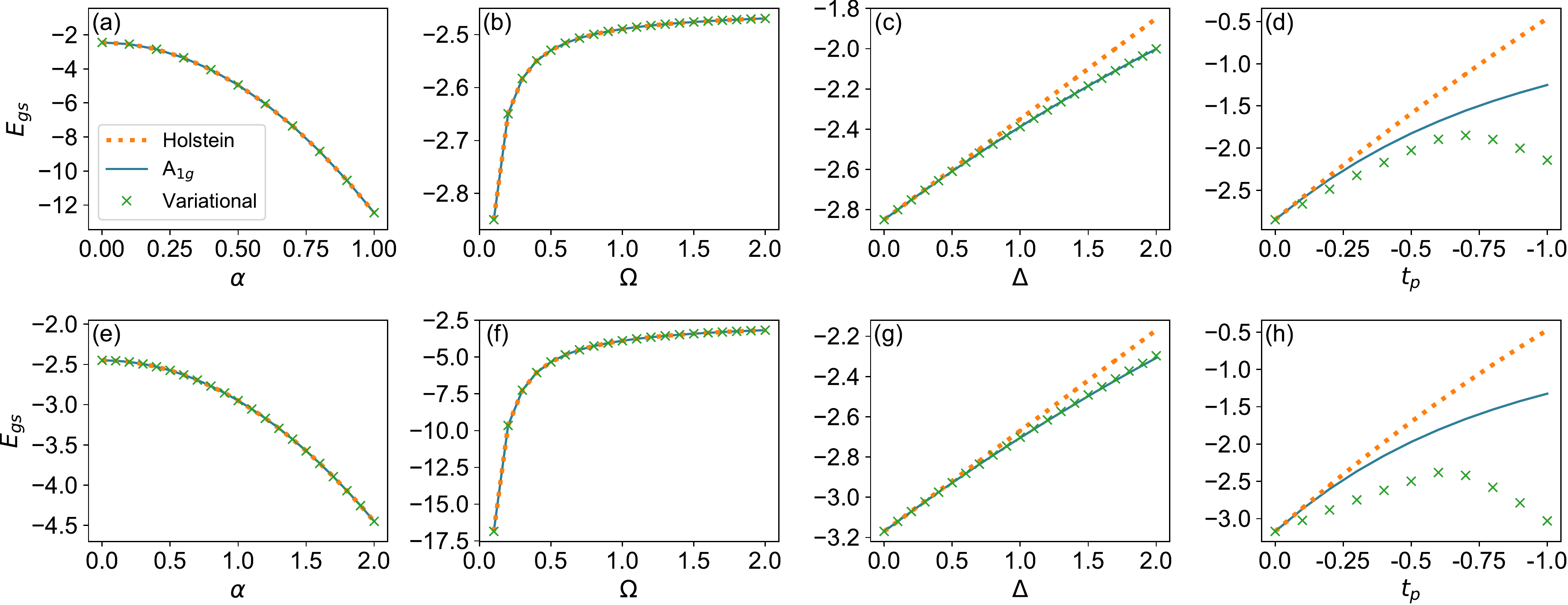}
	\caption{\label{fig3}Comparison between the cluster ground state energies  obtained with the variational (symbols), $A_{1g}$ (full line) and Holstein (dashed line) approximations, respectively, as a function of various parameters. If not otherwise specified, parameters used in (a)-(d) are $t=-1$, $t_p=0$, $\Delta=0$, $\Omega=0.1$ and $\alpha=0.2$; and in  (e)-(h)  $t=-1$, $t_p=0$, $\Delta=0$, $\Omega=2$ and $\alpha=1.2$. In all cases,  $\beta=\frac{\alpha}{2}$.}
\end{figure*}

\subsection{Holstein approximation for the cluster Hamiltonian}

We rewrite the electronic part of ${\cal H}_{A_{1g}}$ in terms of the (for holes) bonding  $d_1^{\dagger}=(s^{\dagger}-P_1^{\dagger})/\sqrt{2}$ and anti-bonding $d_2^{\dagger}=(s^{\dagger}+P_1^{\dagger})/\sqrt{2}$ operators. For $T\gg \Delta $, the bonding orbital is located about $2T$  below the anti-bonding one, and we expect to get a good low-energy approximation by ignoring all terms involving $d_2$ operators. 

The resulting simplified cluster Hamiltonian is:
\begin{align*}
{\cal H}_H =  &\frac{\Delta}{2}d_1^{\dagger}d_1+\Omega B_1^{\dagger}B_1-Td_1^{\dagger}d_1
+\alpha t d_1^{\dagger}d_1(B_1^{\dagger}+B_1)\\
&\qquad -2\left[t_p+\sqrt{\frac{2}{3}}\beta t_p(B_1^{\dagger}+B_1)\right]d_1^{\dagger}d_1\\
= & \epsilon d_1^{\dagger}d_1+\Omega B_1^{\dagger}B_1+
g_Hd_1^{\dagger}d_1(B_1^{\dagger}+B_1)
\end{align*}
This defines a one-site Holstein model with effective parameters $\epsilon = \frac{\Delta}{2}-T-2t_p$ and $g_H= \alpha t-2\sqrt{\frac{2}{3}}\beta t_p$. It can be solved exactly, and has a one-hole ground-state energy $E_H = \epsilon - g_H^2/\Omega$. In the following, we refer to this as the ``Holstein'' cluster energy.

By comparing the variational, $A_{1g}$ and Holstein cluster energies, we can infer the validity of these various approximations in different regions of the parameter space, to see  when/if a Holstein model provides a good low-energy description of the cluster. Together with the results for the lattice case, this will allow us to understand the equivalence (or lack theoreof) between the Peierls and the Holstein models on the perovskite lattice.

\section{Results \label{secIV}}

In this section, the values chosen for the various parameters  are for illustration purposes, so that a broad region of the parameter space can be sampled. Results specific to the values we believe to be appropriate for BaBiO$_3$ are presented and discussed in the last section.

\subsection{Results for the cluster}

Figure \ref{fig3} compares the cluster ground-state energies $E_{gs}$ obtained  with the  variational (symbols), $A_{1g}$ (full line) and Holstein (dashed line) approximations. In all cases, we use $|t|=1$ as the unit of energy. Panels (a) and (e) show the evolution of $E_{gs}$ with the Peierls coupling $\alpha$ (with $\beta=\alpha/2$ used throughout), when $t_p=0$ and $ \Delta=0$. The phonon frequency is $\Omega=0.1$ in panel (a) and $\Omega=2$ in panel (e). The three approximations are in very good agreement.
The same is true for panels (b) and (f), where we track the dependence of $E_{gs}$ on $\Omega$. Here, we continue to keep $t_p=\Delta=0$, and we set the Peierls coupling $\alpha=0.2$ in panel (b) and $\alpha=1.2$ in panel (f). We conclude that for vanishing $\Delta$ and $p$-$p$ hopping, a Holstein description is very satisfactory for the cluster for all phonon frequencies and electron-phonon couplings.

This is no longer the case, however, when either  $\Delta \ne0$ and/or $t_p\ne0$. In panels (c) and (g) we track the dependence of $E_{gs}$ on $\Delta$, when $t_p=0$ and $\Omega=0.1, \alpha=0.2$ in panel (c), versus $\Omega=2, \alpha=1.2$ in panel (g). Both cases show good agreement between the variational and the $A_{1g}$ results, suggesting that the cluster distortion remains s-like. However, projecting out the anti-bonding orbital becomes increasingly inaccurate with increasing $\Delta$. This is because a large $\Delta$ favors a different mix between the $s$ and  $P_1$ orbitals than the $50/50$ mix favored by the hybridization $T$ and by the electron-phonon coupling, see Eq. (\ref{HA1g}). As a result, there is no unique choice for a  single ``cluster" low-energy electronic orbital onto which to project, thus a Holstein-like description becomes increasingly inaccurate.

The problem is further exacerated if we add a finite $t_p$ hopping. This term is known to be important because it is primarily responsible for setting the bandwidth of the O band, which is generally considerable in perovskites. Physically, this is a consequence of the rather short distance between adjacent O, which means that $t_p$ is not negligible compared to $t$. As already noted, it also decreases the effective charge transfer energy between the $s$ and $P_1$ orbitals. The dependence of $E_{gs}$ on $|t_p|$ is shown in panels (d) and (h). In both cases $\Delta=0$, and the values of the other parameters are as in (c) and (g), respectively. For any finite $t_p$, the $A_{1g}$ approximation fails rather fast, and the  Holstein one is even worse. The reason is that the $\beta$ Peierls coupling connects the $s$-like O distortion described by $B_1, B_1^\dagger$ not just to the  $P_1$ orbital with $A_{1g}$ symmetry, but also to the $E_g$ orbitals $P_2, P_3$, see Eq. (\ref{H_sym}). In term, when the electron occupies one of these other orbitals, it favors the appearance of distortions with the same symmetry, see the $\alpha$ term in Eq. (\ref{H_sym}). The end result is that the other distortion modes are also activated, so now even the projection onto  the $A_{1g}$ symmetry is inaccurate, making the further steps to a Holstein mapping impossible.

\begin{figure}[t]
	\centering
	\includegraphics[width=0.65\columnwidth]{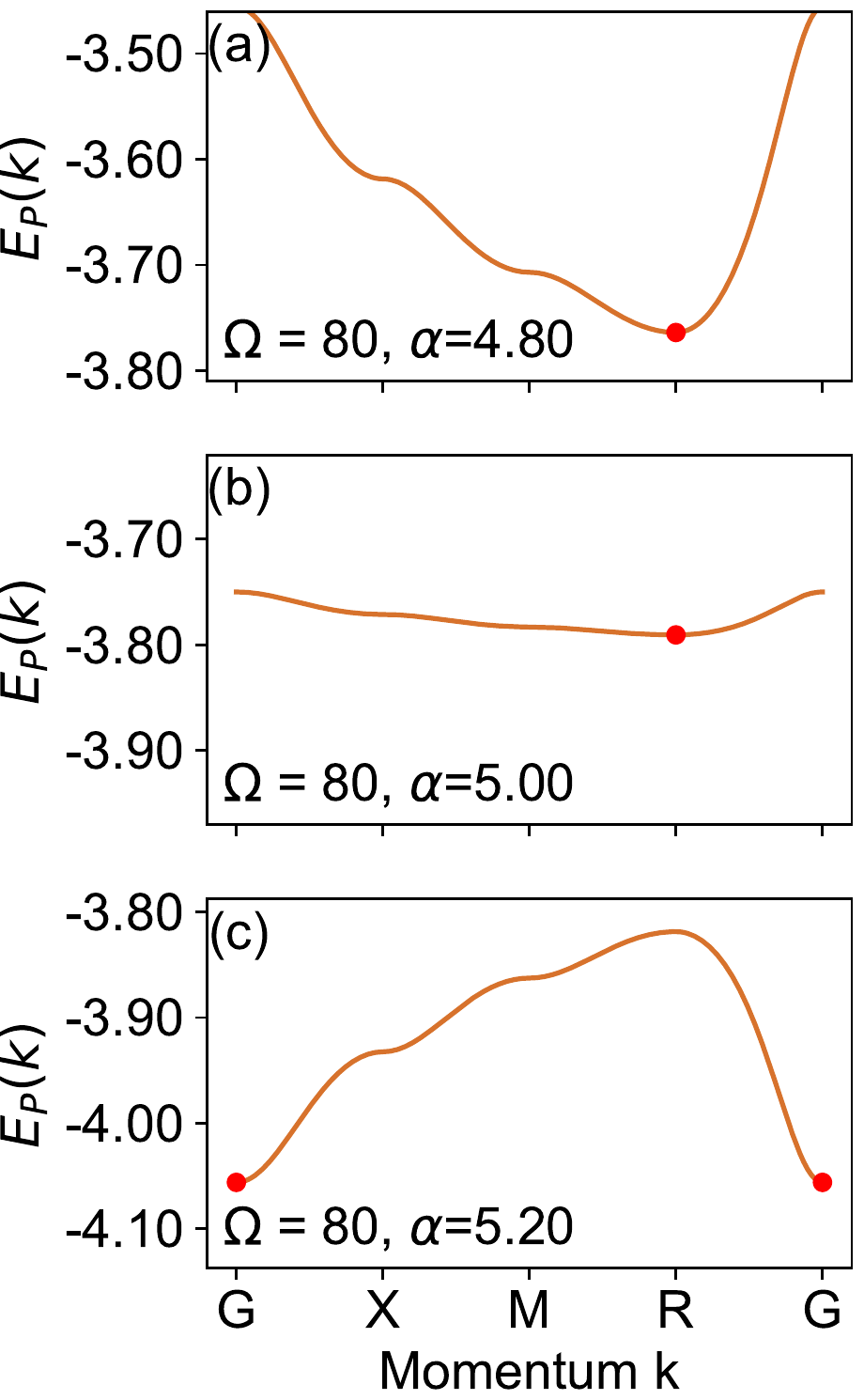}
	\caption{\label{fig4} Polaron dispersion in the anti-adiabatic limit. These are perturbational results but in excellent agreement with the MA  results. Parameters are $|t|=1, t_p=\Delta=0, \Omega=80$ and (a) $\alpha=4.8$, (b) $\alpha= 5$, and (c) $\alpha=5.2$. Red dots indicate the ground-state.}
\end{figure}

Indeed, we find that the downturn of the variational energy at larger  $t_p$ occurs because the $E_{g}$ symmetry starts to dominate over the $A_{1g}$ one, as shown by their weights in the variational calculation (not shown here). This is reminiscent of the phase transition\cite{arash2018} in BaBiO$_3$
  between the $A_{1g}$ dominated bond-disproportionated state and the $E_{g}$ metallic state, driven by the change of effective charge transfer energy $\Delta_1=\Delta+4t_p$. 

It is important to emphasize that the activation of the cluster bosonic modes with other $E_g$ symmetries does not necessarily imply a non-symmetric distortion of the O cage  ({\it i.e.} one breaking the cubic symmetry), so far as the average distortion is concerned. For example, activation of the  $B_3$ ($x^2-y^2$)  distortion will either bring  the O on the $x$-bonds closer and push the $y$-bonds O further out, or viceversa. A wavefunction which has equal contributions from both positive and negative $B_3$ distortions will, in average, retain the cubic symmetry. 

To conclude, the cluster results already demonstrate that an effective Holstein description is likely to fail for realistic systems with finite charge-transfer energies $\Delta \ne0$ and finite $p$-$p$-hopping $t_p \ne 0$.

\subsection{Results for the lattice}

\begin{figure}[b]
	\centering
	\includegraphics[width=1\columnwidth]{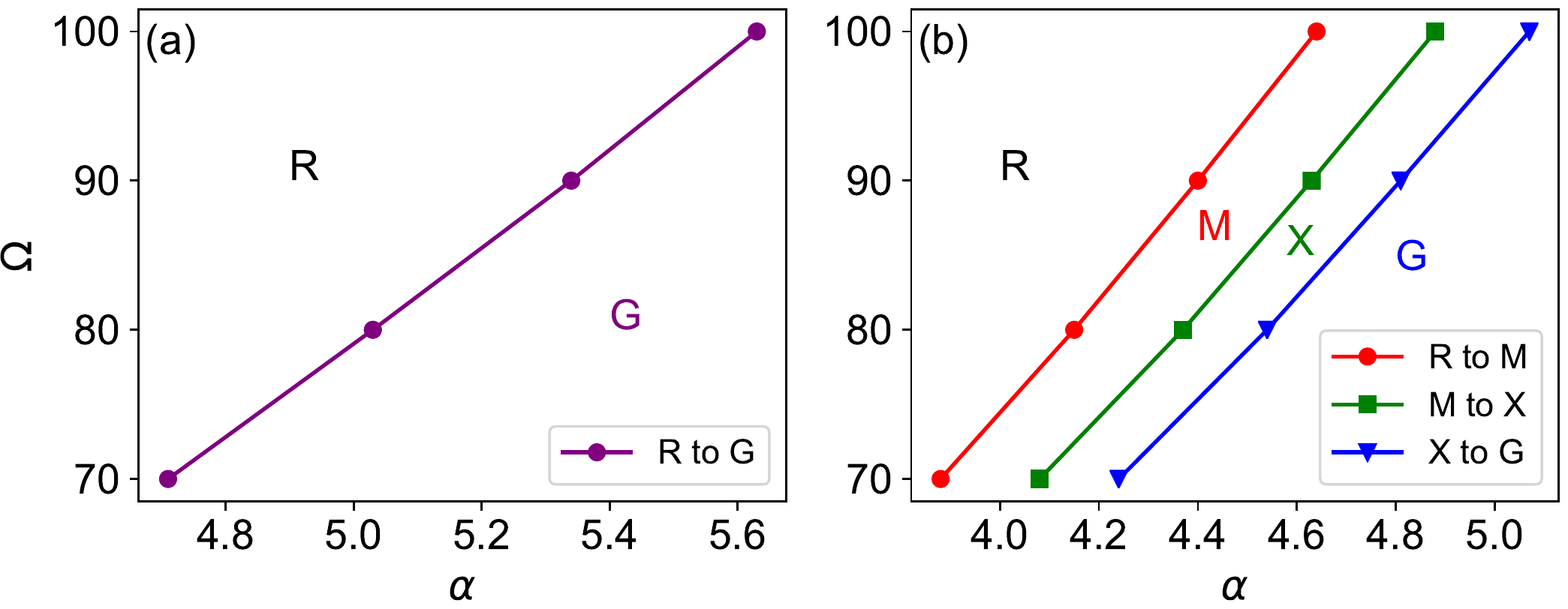}
	\caption{\label{fig6} (a) Ground state momentum $\bm{k}_{gs}$ in the ($\Omega$, $\alpha$) parameter space, showing a sharp transition from $\bm{k}_{gs}=R$ to $\bm{k}_{gs}=G$. Here, $|t|=1, t_p=0, \Delta=0$. (b) same as in (a) but for $t_p=-0.2$. In this case, $\bm{k}_{gs}$ moves from $R\rightarrow M \rightarrow X \rightarrow G$. }
\end{figure}

As just shown, the cluster results indicate that the Holstein mapping is not valid in parts of the parameter space. We expect this conclusion to be even stronger for the lattice case, given its lower symmetry group.

To gain some intuition, we first use perturbation theory to study the evolution of the polaron dispersion in the anti-adiabatic limit $\Omega \gg t$. Of course, this limit is not physical, {\it i.e.} most materials are rather in the adiabatic limit (although this may change for ``flat-band'' materials). However, as we show  below, the qualitative behaviour remains similar for all values of $\Omega$.

We first set $\Delta=0, t_p=0$ and  study the evolution of the polaron dispersion with increasing  $\alpha=4.8, 5, 5.2$  (and $\beta = \alpha/2$) for $\Omega=80$, $|t|=1$. The results are shown in Fig. \ref{fig4}. As customary, the high-symmetry points in the cubic Brillouin zone are $G=(0,0,0)$, $M=(\pi,\pi,0)$, $X=(\pi,0,0)$ and $R=(\pi,\pi,\pi)$ (we set $a=1$).

For a Holstein model, the polaron dispersion has roughly the same shape as the free-hole band, but its bandwidth decreases monotonically with increasing electron-phonon coupling. In contrast, here we see that for $\alpha > 5$, the bandwidth starts to increase again. This is associated with a sharp switch of the momentum of the ground-state from $R$ to $G$, {\it i.e.} a change of the shape of the dispersion that is impossible  for a Holstein model \cite{gerlach91}. In Fig. \ref{fig6}(a) we show the location of this sharp transition in the  $(\alpha, \Omega)$ plane, when $t_p=0, \Delta=0$ and $\Omega \gg1$.

\begin{figure}[t]
	\centering
	\includegraphics[width=0.6\columnwidth]{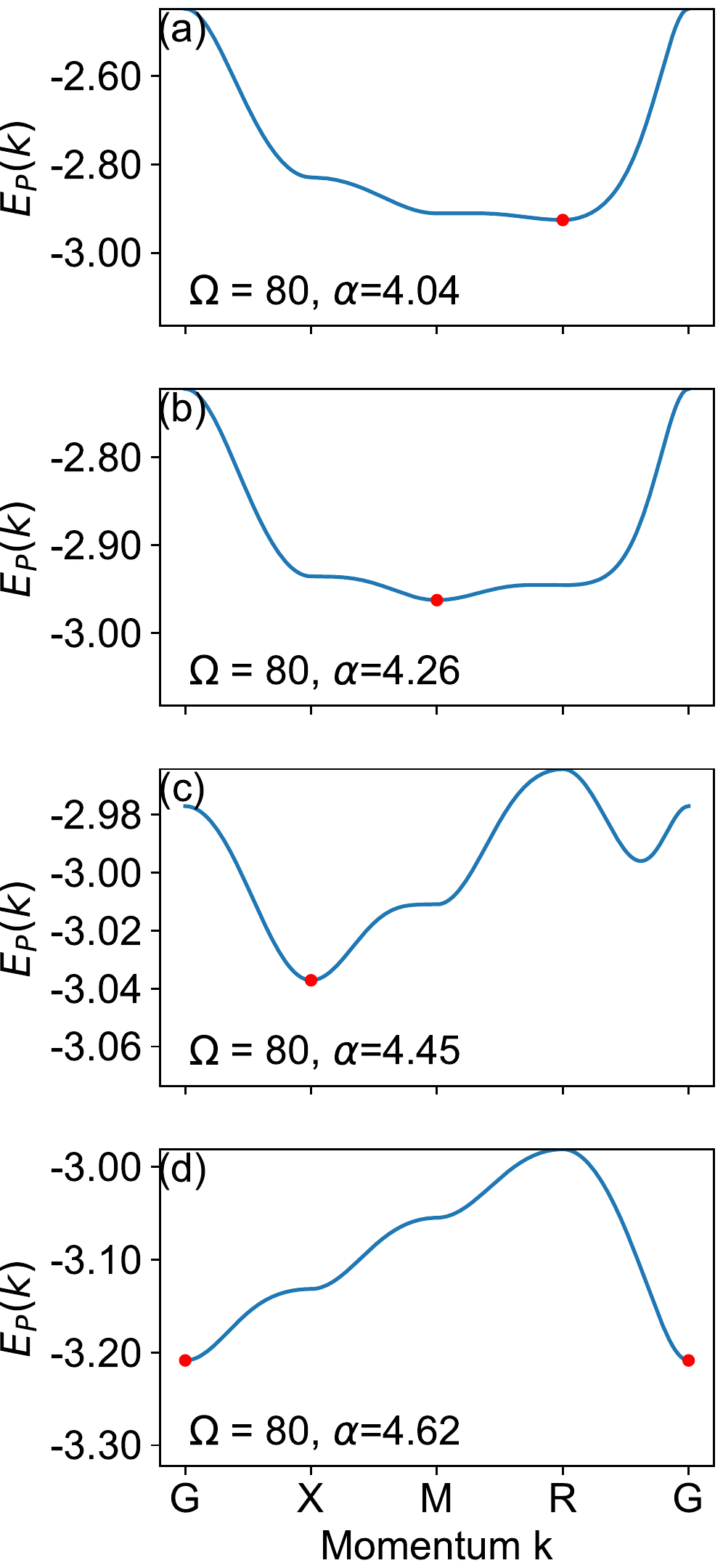}
	\caption{\label{fig7} Evolution of the polaron dispersion with $\alpha$, when $t_p=-0.2$. Parameters not explicitly listed in the panels are as for Fig. \ref{fig4}.}
\end{figure}

We now consider what happens when $\Delta\ne 0, t_p \ne 0$. We find that if $t_p=0$, setting $\Delta\ne 0$ simply shifts the location of the transition in the parameter space (not shown). More spectacular  is the case  $t_p\ne0$, where  as $\alpha$ increases, we find not one but three closely spaced ground-state transitions from  $R \rightarrow M \rightarrow X \rightarrow G$, see Fig. \ref{fig6}(b). The evolution of the polaron dispersion across these transitions is shown in Fig. \ref{fig7}, again for an unphysical value $\Omega=80$.

\begin{figure}[t]
	\centering
	\includegraphics[width=0.7\columnwidth]{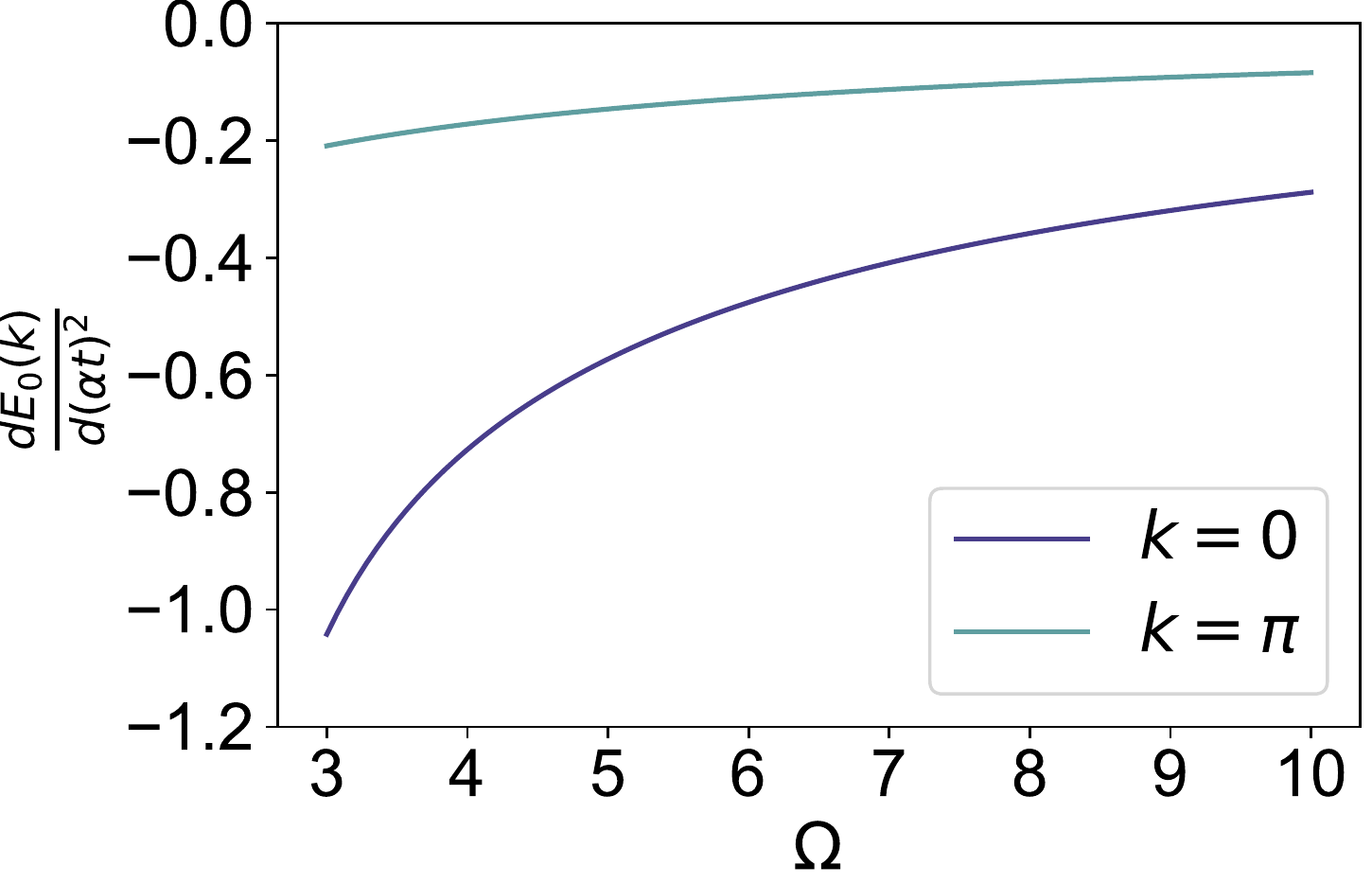}
	\caption{\label{fig5} Measure of the polaron energy $E_P(k)$ on the coupling $\alpha$, in the limit $\alpha \rightarrow 0$. Note that for $k=0$, the polaron energy decreases faster with $\alpha$ than for $k=\pi$. These are 1D results obtained with weak-coupling PT of Eq. (\ref{wPT}). Parameters are $t=-1, t_p=0, \Delta=0$. }
\end{figure}

Similar sharp transitions (sudden jumps) of the GS momentum between high-symmetry
points have also been found for the 1D and 2D versions of this model,
see Refs. \onlinecite{moller2016,moller2017}. They can be
understood in several ways.

\begin{figure}[b]
	\centering
	\includegraphics[width=0.9\columnwidth]{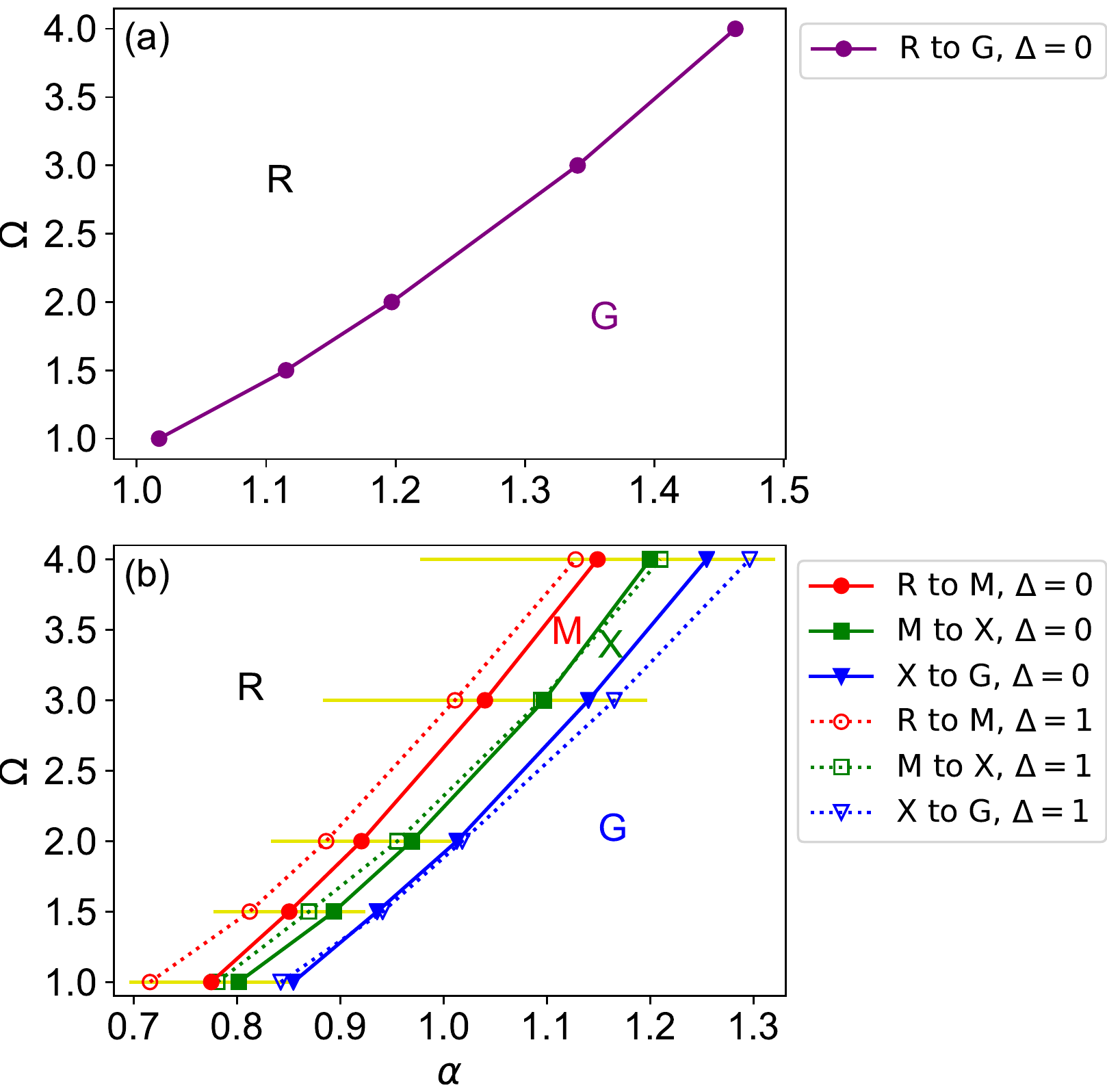}
	\caption{\label{fig8}Ground state momentum transition in $(\alpha,\Omega)$ space, for smaller $\Omega$. Panel (a) is for $t_p=0, \Delta=0$, while panel (b) is for $t_p=-0.2, \Delta=0$ (straight lines) and  $t_p=-0.2, \Delta=1$ (dotted lines). Yellow markers show typical size of errors in locating transitions. In the MA calculations for  these graphs, $50^3$ unit cells are used for the 3D lattice and 300 terms are summed in the Chebyshev expansions for the free-hole propagators $G_0^{\alpha\beta}({\bf{k}},\omega)$. Other parameters are $t=-1$ and $\eta=0.1$.}
\end{figure}

In the anti-adiabatic limit, PT shows that the main effect of the Peierls electron-phonon coupling is to dynamically generate longer-range hopping terms and to renormalize the charge-transfer energy, see $\delta \hat{h}$ of Eq. (\ref{delta_h}) and following discussion. These longer-range hoppings favor a different  $\bm{k}_{gs}$ than that of the bare-hole dispersion, and thus the transition occurs when the electron-phonon coupling is strong enough that these new terms dominate the polaron dispersion. For $t_p\ne 0$, the number of such phonon-mediated longer-range hoppings increases further and the resulting, more complex polaron dispersion, has more transitions.  More discussion along these lines is available in  Ref. \onlinecite{moller2016}.

This argument, however, is predicated on the system being in the anti-adiabatic limit, and thus one might wonder if similar physics is seen at lower, more physical values of $\Omega$. Before showing results proving that this is indeed the case, we first provide a second argument explaining the origin of the $\bm{k}_{gs}$ jump(s). This is based on PT for weak electron-phonon coupling. For simplicity, we carry out this analysis for the 1D equivalent of our 3D model. 

\begin{figure}[t]
	\centering
	\includegraphics[width=0.65\columnwidth]{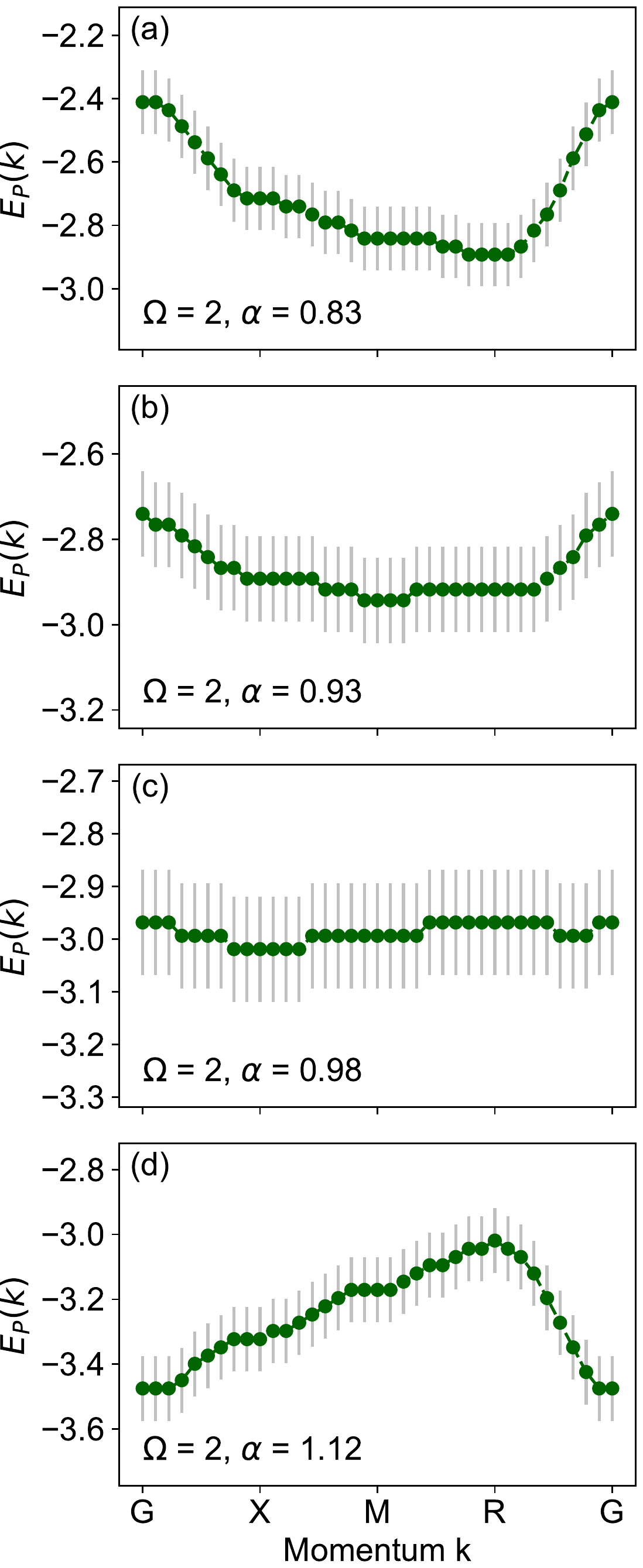}
	\caption{\label{fig9} Polaron dispersion calculated with MA for $t=-1, \Omega=2, t_p=-0.2, \Delta=0$  and $\alpha$ values as indicate on the panels. For small $\alpha$ we find $\bm{k}_{gs}=R$, see panel (a), while for large $\alpha$ we find $\bm{k}_{gs}=G$. At intermediary values, the GS is also found at M and X.}
\end{figure}

\begin{figure*}
	\centering
	\includegraphics[width=0.83\paperwidth]{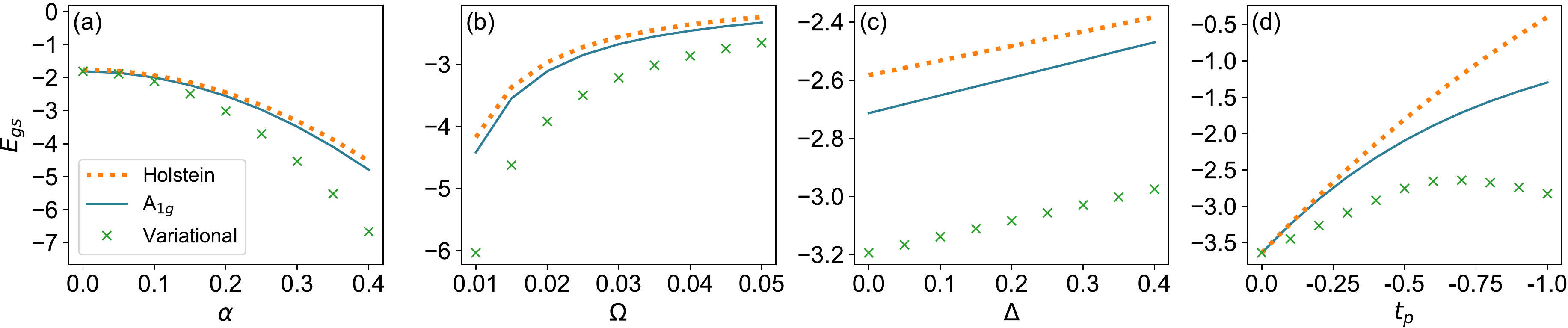}
	\caption{\label{fig10}Comparison between the cluster ground state energies using parameters of BaBiO$_3$ obtained with the variational (symbols), $A_{1g}$ (full line) and Holstein (dashed line) approximations, respectively, as a function of various parameters. If not otherwise specified, parameters used in (a)-(d) are the BaBiO$_3$ parameters in units of $|t|=2.10$,  namely  $t=-1$, $t_p=-0.3$, $\Delta=0.19$, $\Omega=0.033$ and $\alpha=0.206$.}
\end{figure*}

As shown in Eq. (\ref{wPT}), the PT expression for $E_P(k)$ depends on $k$ not just through the usual energy denominators, but also because of the explicit $(k,q)$ dependence of the Peierls electron-phonon vertex. The latter essentially means that holes with different momenta $k$ couple  with different strengths to the phonons, and this will affect how fast their energy is lowered with increasing $\alpha$. Indeed, in Fig. \ref{fig5} we plot $d E_P(k)/ d (\alpha t)^2$ when $\alpha\rightarrow 0$, as a measure of this dependence of $E_P(k)$ on $\alpha$. Both at $k=0$ and at $k=\pi$ the values are negative, as expected, showing a lowering of the energy in the presence of electron-phonon coupling. However, the slopes are very different, with  $E_P(0)$ moving faster towards lower-energies than $E_P(\pi)$. This explains how it is possible that at a large enough $\alpha$, the GS momentum will switch from the free-hole value $k_{gs}=\pi$ to $k_{gs}=0$, instead. Also note that this difference is enhanced as one moves towards the adiabatic  limit, suggesting that the existence of the transition(s) should be expected for any $\Omega$, not just in the anti-adiabatic limit. We confirm this below.

The existence of these sharp transitions of the ground-state of the Peierls model shows that it cannot be universally mapped onto a simpler Holstein model, because the ground-state of the latter cannot exhibit sharp transitions, instead its ground-state momentum is pinned at the free-hole value. Interestingly, the transition appears  even for $t_p=0, \Delta=0$, see Fig. \ref{fig6}(a), so even here the mapping of Peierls onto Holstein does not work for the lattice case,  even though the single cluster is modelled well by a Holstein Hamiltonian. This is proof of the fact that adequate cluster mapping is a necessary but not a sufficient condition for adequate lattice mapping.

Finally, we use MA to show that qualitatively similar behavior is seen at lower, more physical values of $\Omega$. Indeed, we find that the sharp transitions persist, specifically again  if $t_p=0$ there is one from $R \rightarrow{G}$, see Fig. \ref{fig8}(a),  and if $t_p\ne 0$ there are three from $R \rightarrow M \rightarrow X \rightarrow G$, see Fig. \ref{fig8}(b). In panel (b) we also show the slight shift of these transition lines if we set $\Delta=1$. Note the rather large error bars in the location of these transitions, shown in Fig. \ref{fig8}(b). Their origin is the difficulty to accurately calculate the free-hole propagators $G_0^{\alpha\beta}({\bf{k}},\omega)$. In order to smooth out fast oscillations in their $\omega$-dependence --  the finite cutoff in the Chebyshev polynomials expansion  means that we are effectively considering a finite size lattice, thus discretizing the free-hole spectrum --  we are forced to use a rather large value $\eta =0.1$. This sets a limit for our accuracy in identifying the polaron energy, also see Fig. \ref{fig9}, which then translates into the uncertainty in figuring out when the ground-state momentum jumps from one high-symmetry point, to another. However, the  shape and evolution of the spectra are consistent with what we found in the antiadiabatic limit thus giving us confidence that these transitions do occur, in other words that the ground-state momentum switches from R to G as the strength of the electron-phonon coupling is increased.

\section{Discussion \label{secV}}

We used the MA approximation as well as various perturbative limits to study single polaron physics on a perovskite lattice inspired by BaBiO$_3$. The main motivation was to study a multi-band model with Peierls-type of electron-phonon coupling, whereby the motion of ions modulates the value of the hopping integrals between sites, to see if it can be mapped onto a much simpler one-band, Holstein model.

We find clear evidence of sharp transitions in the polaron ground-state properties, something that is proved to be impossible for a Holstein model\cite{gerlach91} (more generally, for any $g(q)$ models including the Rice-Sneddon model). This clearly demonstrates that when considered over the entire parameters space, it is impossible to capture the polaronic physics of the Peierls model with a much simpler Holstein model. The latter is inaccurate not just quantitatively, but qualitatively. Moreover, we showed that it is not enough to study a small cluster to decide this matter, instead one truly needs to study a lattice case. This is because the cluster solution suggested that mapping onto a Holstein model is good when $\Delta=t_p=0$, whereas the lattice results demonstrate that even in this case, a sharp transition occurs with increased electron-phonon coupling.  

Thus, our main conclusion is that serious care is needed in deciding which model to use to describe the electron-phonon coupling in complicated structures such as the perovskites, one should not assume that such details do not matter and that the simplest option is safe.

This being said, we re-emphasize the fact that this conclusion is valid in the insulating limit, where there is a single carrier in the system so that a single polaron forms - this limit can be studied with the well-established  MA approximation, supplemented and reinforced with PT results. Unfortunately, at this time we do not have access to similarly accurate approximations that deal with finite concentrations of carriers, so we cannot make any confident claims about  such systems. The same is true even for the single polaron in  the strongly adiabatic limit, where we know that the variational space used for the MA implemented here is too limited. These questions remain to be studied in future work.

Keeping in mind the caveats mentioned above, we now apply these methods to consider BaBiO$_3$. We use $t=-1$, $t_p=-0.3$, $\Delta=0.19$, $\Omega=0.033$ and $\alpha=0.206$ as reasonable parameters, following the work of Ref. \onlinecite{khazraie2018}. A very rough extrapolation of the curves shown in Fig. \ref{fig8}(b) suggests that this point falls to the left of the transition lines (the GS momentum is still at $R$ like for the free carriers) although probably not by a lot.

This is confirmed if we consider the cluster results, shown in Fig. \ref{fig10}. Each panel has all but one of the parameters fixed at the above-mentioned values, while the varying parameter explores a range around the nominal BaBiO$_3$ value. In all cases we see  discrepancies between the solution for the full cluster Hamiltonian (even when studied variationally) and the Holstein projection. In particular, panel (d) shows that $t_p=-0.3$ falls indeed to the left of the downturn where the   $E_g$  modes start to dominate, in other words the $A_{1g}$ is still the important symmetry here, but not by much.

\begin{figure}
	\centering
	\includegraphics[width=0.75\columnwidth]{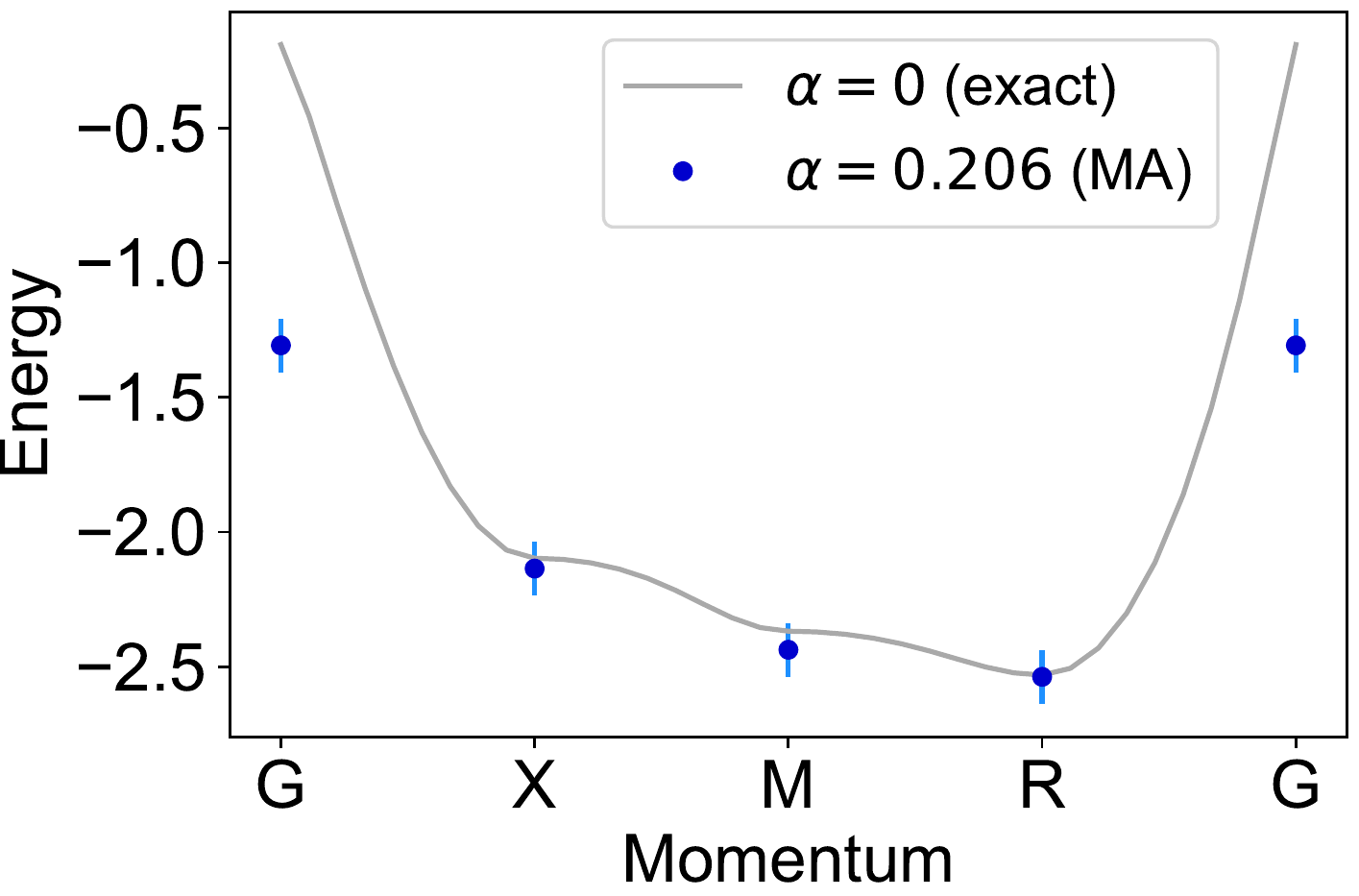}
	\caption{\label{fig11} Dispersion for the 3D lattice Peierls model with parameters of BaBiO$_3$. Blue symbols with error bars show the MA results at the high symmetry points, for a system with 101$^3$ sites  and $\eta=0.1$. For comparison, the dispersion without electron phonon coupling ({\it i.e.} $\alpha=0$) was shown by the gray line, which was solved exactly algebraically. }
\end{figure}

For completeness, in Fig. \ref{fig11} we also show MA results for the polaron dispersion corresponding to these parameters. Given the very low $\Omega/t$ ratio and the large $\eta$ used ($\eta=0.1$), we do not expect these results to be  quantitatively very accurate, however they provide an upper bound on the actual polaron dispersion, because MA is a variational method. The main point is that the GS is still at the $R$ point, {\it i.e.} it is possible that this dispersion could be captured with an appropriately chosen Holstein model. Again, to what extent these conclusions hold for the finite hole concentrations that are physically relevant for BaBiO$_3$, is a matter for future studies. 

\acknowledgements We thank Dr. Lucian Covaci for useful discussions regarding the Chebyshev polynomial expansion. This work was supported by the Steward Blusson Quantum Matter Institute (SBQMI) and by the Natural Sciences and Engineering Research Council of Canada (NSERC).

\appendix

\section{Details of the  MA implementation\label{appMA}}

As discussed in the main  text, we implement the simplest MA$^{(0)}$ version, which allows phonon to appear only at one site in any given configuration. With this restriction, for $\gamma=s,x,y,z$ and $\delta=x,y,z$, we find that
\begin{align*}
f^{(n)}_{\gamma,\delta,\ell}=&(\tilde{f}^{(n+1)}_{\delta,\delta}+n\tilde{f}^{(n-1)}_{\delta,\delta})
[(-\alpha t)(G^{s\gamma}_{0,-\ell}+G^{s\gamma}_{0,-\ell+\delta})\\
&+\beta t_p(\bar{g}^{\delta'\gamma}_{-\ell}+\bar{g}^{\delta''\gamma}_{-\ell})]\\
&+G^{\delta\gamma}_{0,-\ell}[(-\alpha t)(\tilde{f}^{(n+1)}_{s,\delta}+n\tilde{f}^{(n-1)}_{s,\delta})\\
&+\beta t_p(\bar{f}^{(n+1)}_{\delta',\delta}+\bar{f}^{(n+1)}_{\delta'',\delta}+
n\bar{f}^{(n-1)}_{\delta',\delta}+n\bar{f}^{(n-1)}_{\delta'',\delta})]
\end{align*}
where $f \equiv f(\omega)$ while $G_0\equiv G_0(\omega-n\Omega)$ because of the cost of the $n$ phonons present, and then indexes $\gamma', \gamma''$ associated with a given $\gamma$ are defined in the main text following Eq. (3). The free carrier propagators $G_0(\omega)$ are defined as:
\begin{align*}
\left\{\begin{array}{rll}
G^{\gamma\gamma'}_{0,i-j}(\omega)\equiv& \mel{\gamma_i}{\hat{G_0}(\omega)}{\gamma'_{j}}\\
\bar{g}^{\delta'\gamma}_{-\ell}(\omega)\equiv& G^{\delta'\gamma}_{0,-\ell}(\omega)
-G^{\delta'\gamma}_{0,-\ell-\delta'}(\omega)
+G^{\delta'\gamma}_{0,-\ell+\delta}(\omega)\\
&-G^{\delta'\gamma}_{0,-\ell+\delta-\delta'}(\omega)\\
\bar{g}^{\delta''\gamma}_{-\ell}(\omega)\equiv& G^{\delta''\gamma}_{0,-\ell}(\omega)
-G^{\delta''\gamma}_{0,-\ell-\delta''}(\omega)
+G^{\delta''\gamma}_{0,-\ell+\delta}(\omega)\\
&-G^{\delta''\gamma}_{0,-\ell+\delta-\delta''}(\omega)
\end{array}\right. 
\end{align*}
where $i$,$j$,$\ell$ are  site indices. 

Substituting these equations of motion for $f^{(n)}_{\gamma,\delta,\ell}$ into the definitions of 
$\tilde{f}^{(n)}_{s,\gamma}$, 
$\tilde{f}^{(n)}_{\gamma,\gamma}$,  $\bar{f}^{(n)}_{\gamma',\gamma}$ and $\bar{f}^{(n)}_{\gamma'',\gamma}$, we find that the latter define recurrence relations linking propagators with a given   $n$ only to those with  $(n+1)$ and $(n-1)$. In other words, we can define a vector 
$$v_{\gamma,n}^T\equiv \left( \tilde{f}^{(n)}_{s,\gamma} ,  \tilde{f}^{(n)}_{\gamma,\gamma},  \bar{f}^{(n)}_{\gamma',\gamma} , \bar{f}^{(n)}_{\gamma'',\gamma} \right)^T$$
such that the equations of motion can be written in compact form as:
$$v_{\gamma,n}=\alpha_{\gamma n}v_{\gamma,n+1}
+n\alpha_{\gamma n}v_{\gamma,n-1}$$
Here $\alpha_{\gamma n}$ is a known matrix whose entries can be read directly from the equations of motion. Note that for $n=0$, after some simplifications, $v_{\gamma,0}$ can be written in terms of the various propagators $G^{\alpha\beta}(\omega)$ of interest, specifically:
\begin{align*}
v_{\gamma,0}&=\begin{bmatrix}(1+e^{-ik_{\gamma}a})G^{\beta s}(\omega)\\G^{\beta\gamma}(\omega)\\
(1-e^{ik_{\gamma'}a}+e^{-ik_{\gamma}a}-e^{i(k_{\gamma'}-k_{\gamma})a})G^{\beta\gamma'}(\omega)\\
(1-e^{ik_{\gamma''}a}+e^{-ik_{\gamma}a}-e^{i(k_{\gamma''}-k_{\gamma})a})G^{\beta\gamma''}(\omega)\end{bmatrix}
\end{align*}
where $P_{\gamma}$ is a matrix and $\tilde{v_0}$ is defined to be 
$(G^{\beta s},G^{\beta x},G^{\beta y},G^{\beta z})^T$.

Such matrix recurrence relations are solved with the ansatz $v_{\gamma,n}=A_{\gamma,n}v_{\gamma,n-1}$
which allows us to calculate the matrices $A_{\gamma,n}$ recursively, starting from $A_{\gamma,N}=0$ for a sufficiently large $N$. This $N$ defines the largest number of phonons allowed to appear in a self-energy diagram, and is increased until the results converge. Once $A_{\gamma,n=1}$ is known, the various propagators $G^{\alpha\beta}(\omega)$ are obtained from Eq. (8). 

Peaks in the spectral weights $-\frac{1}{\pi}\Im G^{\alpha \beta}(\omega)$ indicate the eigenenergies of $\hat{H}$, and thus allow us to determine the lowest eigenenergy for any given momentum $k$.

\section{Chebyshev Polynomial Expansion for Free Propagators}
\label{CP}

Such expansions are well established for a variety of problems.  Here we briefly summarize the main steps, following Ref. \onlinecite{ferreira2015critical}.

Chebyshev polynomials $T_n(x)\equiv \cos(n\cos^{-1}(x))$ are well defined only for $x\in [-1,1]$, thus we need to rescale the range of eigenvalues of the non-interacting Hamiltonian $H_0$ before applying the Chebyshev expansion to it. $E_{max}$ and $E_{min}$ can be found by Fourier transforming $H_0$ to momentum $k$ space and maximizing or minimizing the energies in the $k$ parameter space. We define $a=\frac{E_{max}-E_{min}}{2}$ and $b=\frac{E_{max}+E_{min}}{2}$, and write the normalized Hamiltonian as $\tilde{H_0}=\frac{H_0-b}{a}$ and denote the corresponding non-interacting Green's function as $\tilde{G_0}(\tilde{\omega})$, where $\tilde{\omega}$ is the scaled energy.

We expand \cite{ferreira2015critical}
\begin{align*}
\tilde{G}^{\alpha\beta}_{0,j}(\tilde{\omega})=
\sum^{\infty}_{n=0}2i^{-1}\frac{(\tilde{\omega}-i\sqrt{1-\tilde{\omega}^2})^n}
{\sqrt{1-\tilde{\omega}^2}}\frac{\mel{\alpha_j}{T_n(\tilde{H_0})}{\beta_0}}{1+\delta_{n0}}
\end{align*}
where $T_0(x)=1$, $T_1(x)=x$ and $T_{n+1}(x)=2xT_n(x)-T_{n-1}(x)$.

If we define $\ket{J_n}\equiv T_n(\tilde{H_0})\ket{\beta_0}$, then
$
\ket{J_{n+1}}=2\tilde{H_0}\ket{J_n}-\ket{J_{n-1}}$, thus these $\ket{J_n}$ can be determined recursively starting from 
$\ket{J_0}=\ket{\beta_0}$ and 
$\ket{J_1}=\tilde{H_0}\ket{\beta_0}$. The summation is truncated at a value large enough so that $\tilde{G}^{\alpha, \beta}_{0,j}$ is converged. We note here that there are unphysical oscillations in the $\tilde{G}^{\alpha, \beta}_{0,j}$ obtained if plotted versus energy. This is caused by the standing waves selected due to the finite size of the system. Since $\eta$ is inversely proportional to the lifetime of the state, we can either increase the size of the system or use a larger $\eta$ so that the state cannot live long enough to reach the edge of the system, and hence the finite-size oscillations are smoothed out. Having a larger system would increase the demand for computational power exponentially, therefore we are forced to use a fairly large $\eta$ (0.1 in our case) in order to get a smooth enough curve for $\tilde{G}^{\alpha, \beta}_{0,j}$.

Rescaling back, the various free propagators are: $G^{\alpha\beta}_{0,j}(\omega+i\eta)=\frac{1}{a}\tilde{G}_{0,j}^{\alpha\beta}
\bigg(\frac{\omega-b}{a}+i\frac{\eta}{a}\bigg)$. 

\section{Details of the continued fraction solution for the cluster \label{AppC}}

We consider the Hamiltonian of Eq. (11), where two different electronic orbitals $s$ and $P_1$ (renamed $p$ in the following, for simplicity) are coupled to the same boson mode $B_1$ (renamed $b$ in the following, for simplicity).  The full Hilbert space corresponding to the one-carrier sector is spanned by the  basis $\{|s,n\rangle\equiv \frac{s^{\dagger}(b^{\dagger})^n\ket{0}}{\sqrt{n!}}, |p,n\rangle\equiv\frac{p^{\dagger}(b^{\dagger})^n\ket{0}}{\sqrt{n!}}\}$ with $n\geqslant0$.

We define the propagators:
\begin{align*}
  \left\{ \begin{array}{rcl}
   \mathcal{S}_n(m,z) & \equiv \langle s,n| \hat{G}(z)|s, m\rangle \\
   \mathcal{P}_n(m,z) & \equiv \langle s,n| \hat{G}(z)|p, m\rangle 
\end{array}\right.
\end{align*}
Their equations of motion are generated from the appropriate expectation values of the identity $\hat{G}(z)(z- \hat{H})=1$. For the Hamiltonian of Eq. (11), we find:
\begin{align*}
&\mathcal{S}_n(m,z)(z-\Delta-m\Omega)+\mathcal{P}_n(m,z)t\\
&-\mathcal{P}_n(m+1,z)\alpha t\sqrt{m+1}
-\mathcal{P}_n(m-1,z)\alpha t\sqrt{m}=\delta_{mn}
\end{align*}
and
\begin{align*}
&\mathcal{S}_n(m,z)t-\mathcal{S}_n(m+1,z)\alpha t\sqrt{m+1}\\
&-\mathcal{S}_n(m-1,z)\alpha t\sqrt{m}+\mathcal{P}_n(m,z)(z-\Omega m)=0
\end{align*}
These can be grouped as recurrence equations for $2\times2$ matrices:
\begin{equation*}
\gamma_m W_{nm}-\alpha_m W_{n,m+1}-\beta_m W_{n,m-1}=\begin{bmatrix} \delta_{n,m}\\0 \end{bmatrix} \quad\text{and}
\end{equation*}

where 
\begin{align*}
&\gamma_m\equiv \left[ \begin{array}{cc}z-\Delta-m\Omega & t \\ t & z-\Omega m\end{array} \right]; \\
&\alpha_m\equiv \left[ \begin{array}{cc}0 & \alpha t\sqrt{m+1} \\ \alpha t\sqrt{m+1} & 0\end{array} \right]; \quad \\
&\beta_m\equiv \left[ \begin{array}{cc}0 & \alpha t\sqrt{m} \\ \alpha t \sqrt{m} & 0\end{array} \right]; \\
&W_{n,m}\equiv\begin{bmatrix}\mathcal{S}_n(m,z)\\ \mathcal{P}_n(m,z)  \end{bmatrix}
\end{align*}
As already discussed, such recurrence relations are solved with the ansatz  $W_{n,m+1}=A_{n,m+1}W_{nm}$ if $m\ge n$. This gives the continued fraction  $A_{nm}=(\gamma_m-\alpha_m A_{n,m+1})^{-1}\beta_m$, which  can be evaluated starting from  $A_{nM}=0$ for a sufficiently large $M$. Similarly, for $m \le n$ we  use the ansatz $W_{n,m-1}=B_{n,m-1}W_{n,m}$ and obtain $B_{n,m}=(\gamma_m-\beta_{m}B_{n,m-1})^{-1}\alpha_m$, which can  be computed starting from $m=0$, noting that $\beta_0\equiv 0$.

Putting $W_{n,n+1}=A_{n,n+1}W_{nn}$ and $W_{n,n-1}=B_{n,n-1}W_{n,n}$ into the equation with $n=m$, we get:
\begin{align*}
W_{nn}=(\gamma_n-\alpha_n A_{n,n+1}-\beta_n B_{n,n-1}
)^{-1}\begin{bmatrix}1\\0\end{bmatrix}
\end{align*}
from which we can read out the propagators $\mathcal{S}_n(m,z)$ and $\mathcal{P}_n(m,z)$.

\bibliography{eqv}

\end{document}